\documentclass{aa}
\usepackage{txfonts}
\usepackage[dvips]{color}
\usepackage{graphicx}
\usepackage{natbib}
\bibpunct{(}{)}{;}{a}{}{,}
\usepackage{captcont}
\usepackage{subfigure}
\usepackage{hyperref}
\usepackage{sidecap}

\begin{document}

\title{Resolved gas cavities in transitional disks inferred from CO isotopologs with ALMA}
\titlerunning{Gas cavities in transitional disks}
\author{N. van der Marel\inst{1} 
\and E.F. van Dishoeck\inst{1,2}
\and S. Bruderer\inst{2}
\and S.M. Andrews\inst{3}
\and K.M. Pontoppidan\inst{4}
\and G.J. Herczeg\inst{5}
\and T. van Kempen\inst{1}
\and A. Miotello\inst{1}
}
\institute{Leiden Observatory, Leiden University, P.O. Box 9513, 2300 RA Leiden, the Netherlands
\and Max-Planck-Institut f\"{u}r Extraterrestrische Physik, Giessenbachstrasse 1, 85748 Garching, Germany
\and Harvard-Smithsonian Center for Astrophysics, 60 Garden Street, Cambridge, MA 02138, USA
\and Space Telescope Science Institute, 3700 San Martin Drive, Baltimore, MD 21218, USA
\and Kavli Institute for Astronomy and Astrophysics, Peking University, Yi He Yuan Lu 5, Haidijan district, Beijing 100871, China
}
\date{}
       
\abstract{Transitional disks around young stars with large dust cavities are promising candidates to look for recently formed, embedded planets. Models of planet-disk interaction predict that young planets clear a gap in the gas while trapping dust at larger radii. Other physical mechanisms might also be responsible for cavities. Previous observations have revealed that gas is still present inside these cavities, but the spatial distribution of this gas remains uncertain.}
  {We present high spatial resolution observations with the Atacama Large Millimeter/submillimeter Array (ALMA) of $^{13}$CO and C$^{18}$O 3--2 or 6--5 lines of four well-studied transitional disks around pre-main-sequence stars with large dust cavities. The line and continuum observations are used to set constraints on the the gas surface density, specifically on the cavity size and density drop inside the cavity.}
{The physical-chemical model DALI was used to analyze the gas images of SR21, HD135344B (also known as SAO~206462), DoAr44, and IRS 48. The main parameters of interest are the size, depth and shape of the gas cavity in each of the disks. CO isotope-selective photodissociation is included to properly constrain the surface density in the outer disk from C$^{18}$O emission.}
{The gas cavities are up to three times smaller than those of the dust in all four disks. Model fits indicate that the surface density inside the gas cavities decreases by a factor of 100 to 10000 compared with the surface density profile derived from the outer disk. The data can be fit by either introducing one or two drops in the gas surface density or a surface density profile that increases with radius inside the cavity. A comparison with an analytical model of gap depths by planet-disk interaction shows that the disk viscosities are most likely low, between between 10$^{-3}$ and 10$^{-4}$ , for reasonable estimates of planet masses of up to 10 Jupiter masses.}
{The resolved measurements of the gas and dust in transition disk cavities support the predictions of models that describe how planet-disk interactions sculpt gas disk structures and influence the evolution of dust grains. These observed structures strongly suggest the presence of giant planetary companions in transition disk cavities, although at smaller orbital radii than is typically indicated from the dust cavity radii alone.} 


\maketitle

\section{Introduction}
Protoplanetary disks around young stars are the birth cradles of planets \citep[e.g.,][]{WilliamsCieza2011}. Disks with inner dust cavities, also called transition disks, are good candidates to search for young planets that have recently been formed and cleared out their orbit. Dust cavities have been inferred from modeling of spectral energy distributions (SEDs) \citep[][and references therein]{Espaillat2014} and millimeter interferometry \citep[e.g.,][]{Brown2009,Andrews2011}. Planet candidates have been found in cavities of several transition disks \citep{KrausIreland2012,Quanz2013,Quanz2014}. However, planet-disk interaction models indicate that dust cavities are only an indirect consequence of planet clearing: a planet will lower the gas surface density along its orbit, creating a gas gap with a pressure bump at its outer edge where the millimeter-sized dust is trapped \citep[e.g.,][]{Zhu2011,Dodson-Robinson2011,Pinilla2012b}. The result is a millimeter-dust ring (or in particular cases an azimuthal asymmetry that is due to a Rossby-wave instability of the pressure bump, \citealt{vanderMarel2013,Fukagawa2013, Casassus2013}) and a gas cavity that can be up to twice as small
as the radius of the dust ring. Other mechanisms that could cause a dust cavity or dust ring are photoevaporation \citep{Clarke2001} and instabilities at the edges of dead zones \citep[e.g.,][]{Regaly2012}. Measuring the gas density inside the cavity is essential to distinguish between these mechanisms: photoevaporation clears the dust and gas from the inside out, dead zones do not change the gas density inside the cavity, and planetary clearing creates a gas cavity. Furthermore, the decrease of gas surface density inside the gas cavity radius depends on the mass of the companion and the disk viscosity. More generally, the fundamental properties of a disk are the gas-to-dust ratio and gas surface density profile because the gas does not necessarily follow the dust distribution.

The presence of gas inside the dust cavities was discovered through accretion \citep[e.g.,][]{Valenti1993,Najita2007} and H$_2$ emission \citep{Bergin2003,Ingleby2009}. Other first indicators of warm molecular gas inside the cavities were near-infrared observations of CO rovibrational lines, in several cases revealing a gas cavity that was indeed smaller than the dust cavity \citep{Pontoppidan2008,Brittain2009,Salyk2009,Brown2012a}. As a result of a combination of high critical densities and non-LTE excitation, rovibrational CO data can be difficult to interpret, and derived gas masses are highly model-dependent. 

To properly derive the molecular gas densities, CO pure rotational line observations are required. For a handful of disks, pioneering interferometers such as SMA and PdBI have revealed gas inside the cavity through CO rotational lines: AB~Aur \citep{Pietu2005}, GM~ Aur \citep{Dutrey2008}, and HD~135344B \citep{Lyo2011}. Spatially resolved ALMA observations of $^{12}$CO emission confirm the presence of gas inside the dust cavity for several other disks \citep{vanderMarel2013,Casassus2013,Perez2014,Zhang2014,Canovas2015}. Using a disk model based on the dust structure derived from the SED and millimeter imaging, the $^{12}$CO data suggest a gas density decrease of one or two orders of magnitude inside the dust cavity compared to the density profile of the outer disk \citep{vanderMarel2015-12co}. On the other hand, the dust density decreased by at least three orders of magnitude inside the cavity. For IRS~48 and J1604-2130, the gas cavities are sufficiently empty so that $^{12}$CO becomes optically thin inside the dust cavity, and it was found that the gas cavity radius is indeed smaller than the dust cavity \citep{Bruderer2014,Zhang2014,vanderMarel2015-12co}, consistent with predictions for the planet-disk interaction. The same result was found for HD~142527 using CO isotopolog observations \citep{SPerez2015}. 

Since detecting planets in disks is challenging, quantifying the gas density structure of the disk inside the cavity can provide important clues on the properties of embedded unseen planets. The depth and shape of the gap depend primarily on the planet mass and the disk viscosity \citep{Zhu2011,Pinilla2012b,Fung2014}. These models show that a planet does not create a steep gas gap, but a gradual decrease over several AUs. While spatially resolved $^{12}$CO can provide some information on the gas density profile,  
the emission remains optically thick throughout most of the disk and is thus not a very good absolute density tracer. Spatially resolved CO isotopolog observations are required to constrain the outer disk mass and the gas cavity radius, as well as the depth and potentially the shape of the gas surface density profile. 

Converting CO emission into density is not trivial either: the CO abundance with respect to H$_2$ varies throughout the disk due to photodissociation by the stellar UV radiation and freeze-out onto dust grains in the cold regions in the midplane and outer disk \citep{Zadelhoff2001,Aikawa2002}. CO photodissociation is subject to self-shielding. As CO isotopologs have lower abundances, they are not self-shielded until deeper into a cloud or disk \citep{Bally1982,vanDishoeckBlack1988,Visser2009}. Therefore, disk models that do not include isotope-selective photodissociation predict higher CO abundances than when this effect is included, as recently demonstrated by \citet{Miotello2014}. In addition, the gas temperature is decoupled from the dust temperature in the upper layers in the disk and at the directly heated cavity wall \citep[e.g.,][]{KampDullemond2004,Jonkheid2004,GortiHollenbach2008}. For a proper interpretation of CO emission, the physical and chemical structure of gas and dust needs to be modeled. We make use of physical-chemical modeling with DALI \citep{Bruderer2012,Bruderer2013}, which simultaneously solves the heating-cooling balance of the gas and chemistry to determine the gas temperature, molecular abundances, and molecular excitation for a given density structure. 

\begin{figure*}[!ht]
\begin{center}
\subfigure{\includegraphics[scale=0.4,trim=0 20 0 20]{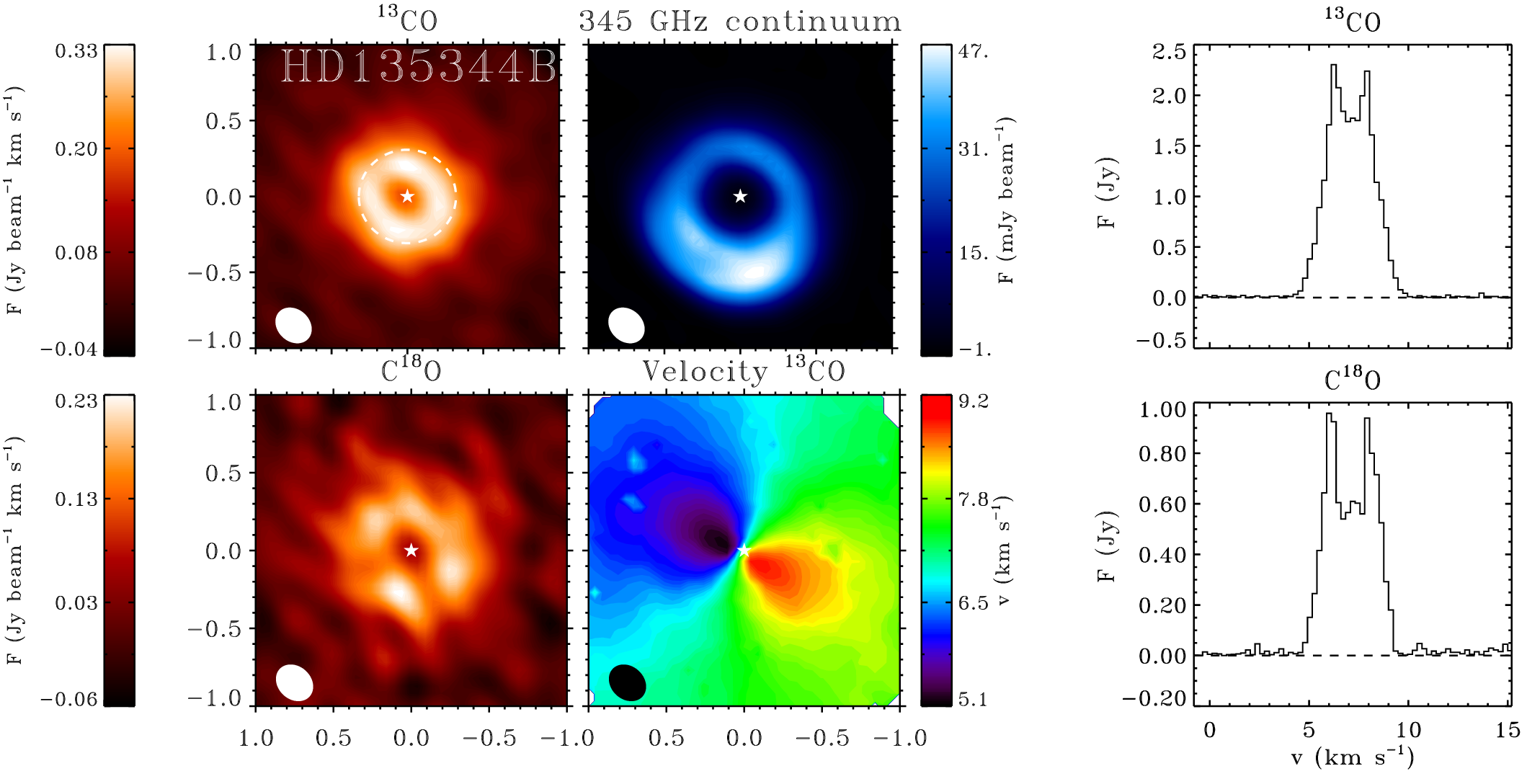}}\\
\subfigure{\includegraphics[scale=0.4,trim=0 20 0 20]{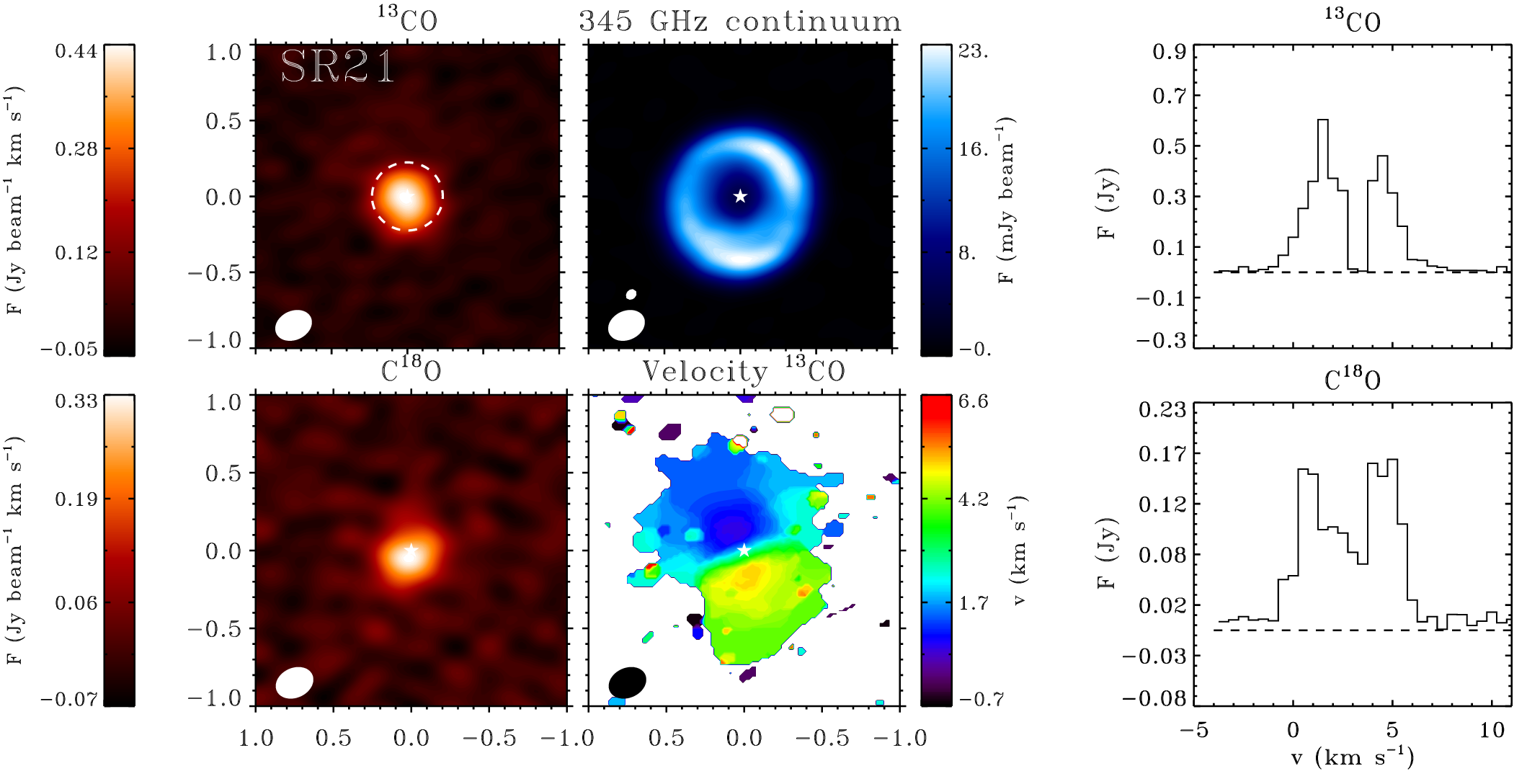}}\\
\subfigure{\includegraphics[scale=0.4,trim=0 20 0 20]{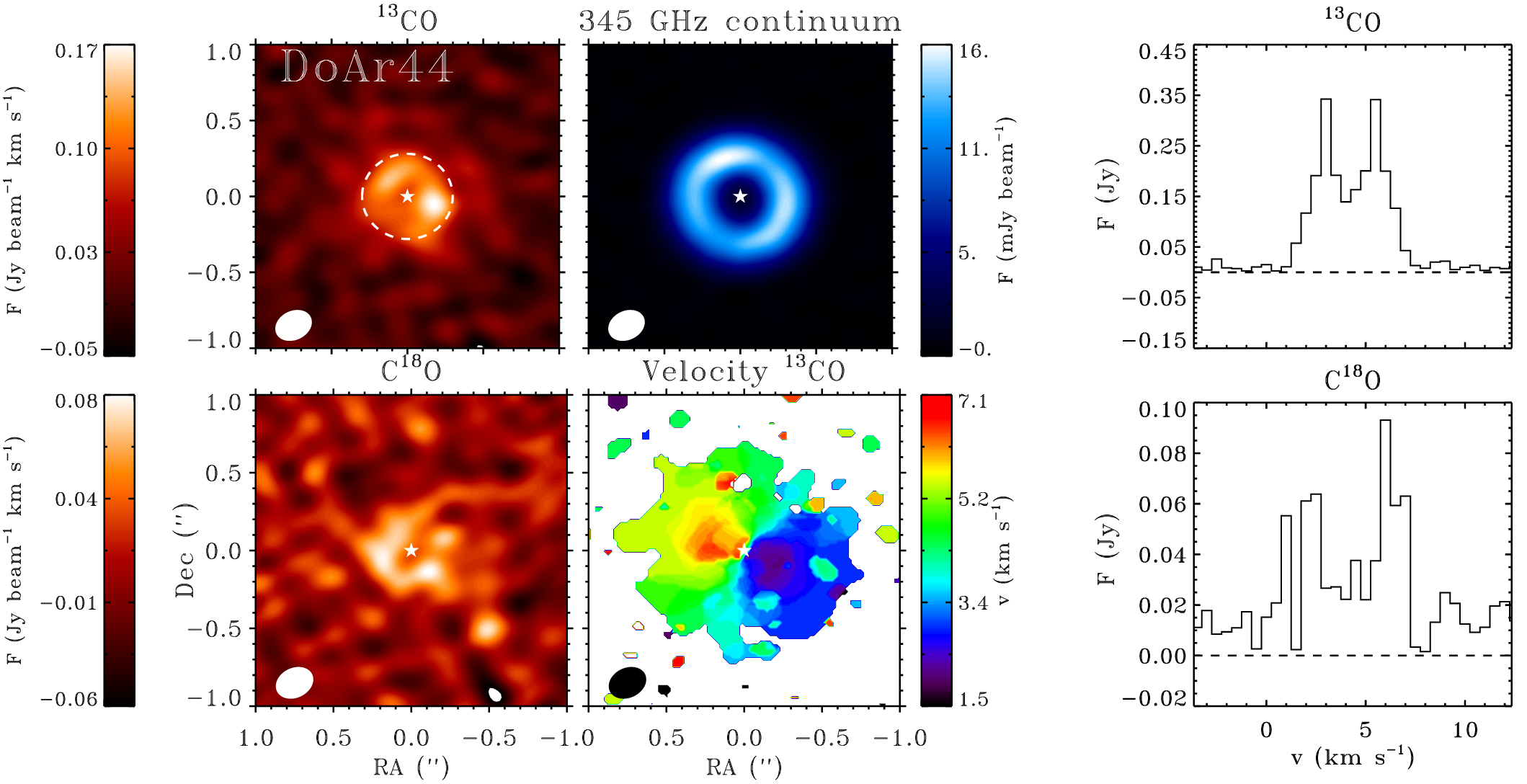}}
\end{center}
\caption{ALMA observations of the continuum, $^{13}$CO and C$^{18}$O 3-2 lines of the first three  targets. {\bf Top left:} zero-moment $^{13}$CO map. {\bf Top middle:} Continuum map. {\bf Top right:} $^{13}$CO spectrum integrated over the entire disk. {\bf Bottom left:} zero-moment C$^{18}$O map. {\bf Bottom middle:} first moment $^{13}$CO map (velocity map). {\bf Bottom right:} C$^{18}$O spectrum integrated over the entire disk. The beam is indicated in each map by a white ellipse in the lower left corner. The dotted white ellipse indicates the dust cavity radius.}
\label{fig:intensity}
\end{figure*}

\begin{figure*}[!ht]
\begin{center}
\includegraphics[scale=0.4,trim=0 20 0 20]{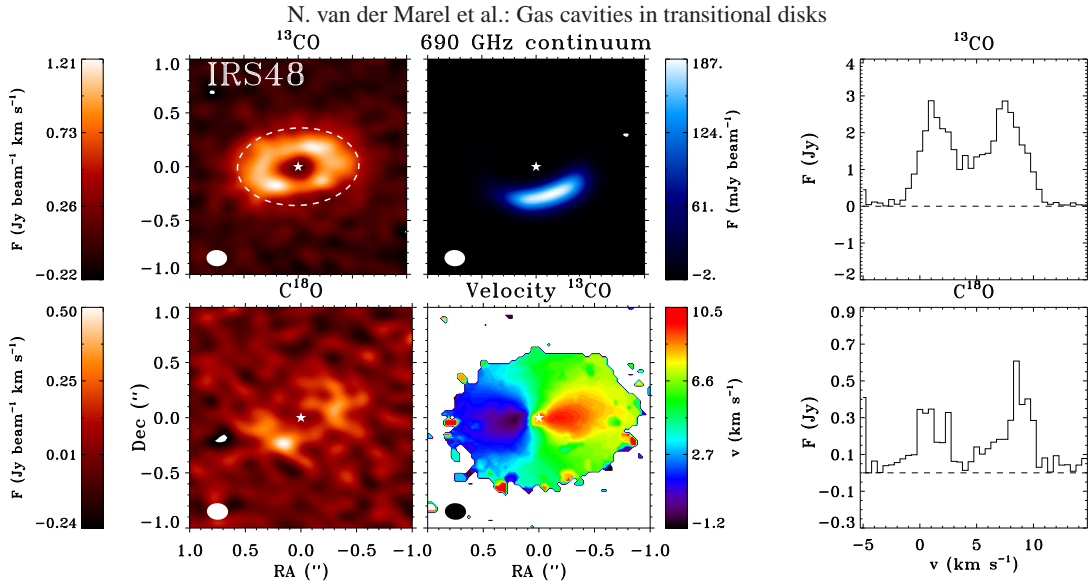}
\end{center}
\caption{ALMA observations of the continuum, the $^{13}$CO and C$^{18}$O 6-5 lines of the fourth target. {\bf Top left:} zero-moment $^{13}$CO map. {\bf Top middle:} Continuum map. {\bf Top right:} $^{13}$CO spectrum integrated over the entire disk. {\bf Bottom left:} zero-moment C$^{18}$O map. {\bf Bottom middle:} first moment $^{13}$CO map (velocity map). {\bf Bottom right:} C$^{18}$O spectrum integrated over the entire disk. The beam is indicated in each map by a white ellipse in the lower left corner. The dotted white ellipse indicates the dust cavity radius.}
\label{fig:intensity2}
\end{figure*}

\begin{table*}[!ht]\small
\begin{center}
\caption{Stellar properties}
\label{tbl:stellar}
\begin{tabular}{llllllllll}
\hline
\hline
Target&SpT&$L_*$&$M_*$&$R_*$&$T_{\rm eff}$&$\dot{M}$&$d$&$A_V$&Ref.\\
&&($L_{\odot}$)&($M_{\odot}$)&($R_{\odot}$)&(K)&($M_{\odot}$ yr$^{-1}$)&(pc)&(mag)&\\
\hline
HD135344B&F4&7.8&1.6&2.2&6590&$6\cdot10^{-9}$&140&0.3&1,2,3\\
SR21&G3&10&1.0&3.2&5830&$<1\cdot10^{-9}$&120&6.3&4,5,6\\ 
DoAr44&K3&1.4&1.3&1.75&4730&$9\cdot10^{-9}$&120&2.2&5,6\\
IRS48&A0&14.3&2.0&1.3&10000&$4\cdot10^{-9}$&120&11.5&7,8\\
\hline
\end{tabular}
\\
\begin{footnotesize}
1) \citet{Prato2003}, 2) \citet{Andrews2011}, 3) \citet{Espaillat2010}, 4) \citet{Dunkin1997}, 5) \citet{Pontoppidan2008}, 6) \citet{Grady2009},  7) \citet{Brown2012a}, 8) \citet{Salyk2013}
\end{footnotesize}
\end{center}
\end{table*}

\begin{table*}[!ht]\small
\begin{center}
\caption{Properties of the ALMA observations}
\label{tbl:obsproperties}
\begin{tabular}{llllllllll}
\hline
\hline
Target&Derived position&Beam&Beam&rms$_{\rm 13CO}$\tablefootmark{a}&rms$_{\rm C18O}$\tablefootmark{a}&rms$_{\rm cont}$&PA&$i$\tablefootmark{b}&$\varv_{\rm LSR}$\\
&(J2000)&size (")&PA ($^{\circ}$)&(mJy beam$^{-1}$)&(mJy beam$^{-1}$)&(mJy beam$^{-1}$)&($^{\circ}$)&($^{\circ}$)&(km/s)\\
\hline
HD135344B&15:15:48.42 -37:09:16.36&0.26$\times$0.21&\phantom{-}46&9.8&14&0.26&63&16&7.25\\
SR21&16:27:10.27 -24:19:13.04&0.25$\times$0.19&-65&6.4&8.9&0.12&14&16&3.0\\
DoAr44&16:31:33.46 -24:27:37.53&0.25$\times$0.19&-65&6.8&8.9&0.14&30&20&4.35\\
IRS48&16:27:37.18 -24:30:35.39&0.19$\times$0.15&\phantom{-}87&23&25&0.59&100&50&4.55\\
\hline
\end{tabular}
\tablefoot{
\tablefoottext{a}{Measured in 0.5 km s$^{-1}$ bins}
\tablefoottext{b}{Derived from the $^{13}$CO channel maps.}
}
\end{center}
\end{table*}

In this paper, we present ALMA Cycle 1 and 2 observations of CO isotopologs $^{13}$CO and C$^{18}$O at $\sim0.2-0.25''$ resolution of four additional well-studied transitional disks: SR~21, HD135344B, DoAr44, and IRS~48. For IRS~48, the 6--5 transitions are observed
and for the other disks the lower 3--2 transitions. 
Previously derived models from $^{12}$CO ALMA observations \citep{Bruderer2014,vanderMarel2015-12co} of IRS~48, SR~21 and HD135344B are used as a starting point for analyzing the isotopologs. With DALI we determine a gas density structure that is consistent with the CO observations, SED, and continuum dust interferometry. Information on the hot gas and dust from the literature is included. The goals of this study are to determine the gas surface density profile, specificially the size, depth, and shape of the gas density structure inside the dust cavity,  to constrain the properties of potential embedded planets.

The paper is structured as follows. In Sect. 2.1 we describe the details of the ALMA observations. In Sect. 2.2 we present moment maps of the $^{13}$CO and $^{18}$CO observations. The modeling approach is presented in Sect. 3. Section 4 presents the modeling results. Section 5 discusses the implications for embedded planets in the disk. 

\section{Data}
The observations were obtained during ALMA Cycle 1 and 2 in June and July 2014, with baselines ranging from 20 to 1100 m, probing scales from 0.15 to 8 arcseconds. The sources and their properties are summarized in Table \ref{tbl:obsproperties}.

\subsection{Observational details}

The disks SR~21, HD~135344B, and DoAr~44 were observed in ALMA Cycle 1 program 2012.1.00158.S (PI van Dishoeck) in Band 7 ($\sim$335 GHz or 896 $\mu$m)  with a resulting spatial resolution of 0.2--0.25". 
The observations were taken in four spectral windows of 3840 channels: three windows have a bandwidth of 469 MHz (channel width 122 kHz, equivalent to 0.1 km s$^{-1}$), centered on the $^{13}$CO 3--2, C$^{18}$O 3--2, and CN 3--2 transitions with rest frequencies of 330.58796, 329.33056, and 340.24778 GHz, respectively. The fourth spectral window was centered on 342.15000 GHz with a bandwidth of 1875 MHz (channel width 488 kHz, equivalent to 0.5 km s$^{-1}$) aimed at higher continuum sensitivity. The total continuum bandwidth was $\sim$3.2 GHz. For HD135344B, the flux was calibrated using Ceres, and J1427-4206 was used for calibration of both bandpass and gain. SR~21 and DoAr~44 were observed in one scheduling block, with Titan as flux calibrator, J1517-2422 as bandpass calibrator, and J1625-2527 as gain calibrator. In both objects, the flux calibrator (Ceres resp. Titan) is highly resolved on long baselines. The gain calibration on the flux calibrator was thus performed using a subset of the antennas. The total on-source integration time was 30 minutes each for SR~21 and DoAr~44, and 54 minutes for HD~135344B. 

IRS~48 was observed in ALMA Cycle 2 program 2013.1.00100.S (PI van der Marel) in Band 9 ($\sim$680 GHz or 440 $\mu$m) with a resulting spatial resolution of 0.15--0.2". 
The observations were taken in four spectral windows of 1920 channels: three windows with a bandwidth of 937.5 MHz (channel width 488 kHz or 0.3 km s$^{-1}$), centered on the $^{13}$CO 6--5, C$^{18}$O 6--5 and H$_2$CO 9--8 transitions with rest frequencies of 661.067276, 658.553278 and 674.80978 GHz, respectively. The fourth spectral window was centered on 672 GHz with a bandwidth of 1875 MHz (channel width 977 kHz or 0.5 km s$^{-1}$) aimed at higher continuum sensitivity. The total continuum bandwidth was $\sim$4.7 GHz. The flux was calibrated using J1517-243, the bandpass with J1427-4206 and the gain with J1626-2951. J1700-2610 was set as secondary gain calibrator, but not used in the final calibration. The total on-source integration time was 52 minutes.   

The data were calibrated and imaged in CASA version 4.2.1. Given the high signal-to-noise ratio of these observations, amplitude and phase self-calibration was performed after standard phase referencing. The data were cleaned using Briggs weighting with a robust factor of 0.5, resulting in a beam size of $\sim0.25"\times0.20"$ (Cycle 1 data) and $\sim0.19"\times0.15"$ (Cycle 2 data). Table \ref{tbl:obsproperties} lists the observational properties of the continuum and spectral line maps of the imaging results.

\subsection{Continuum and line maps}
Figures \ref{fig:intensity} and \ref{fig:intensity2} show the zero-moment $^{13}$CO and C$^{18}$O maps and spectra and the velocity map (first moment) of the $^{13}$CO and the continuum. The spectrum was extracted from the region of the zero-moment map size. Channel maps of the $^{13}$CO emission are given in the Appendix. Three of the four sources, HD135344B, DoAr44, and IRS48, show a clear hole in the $^{13}$CO and C$^{18}$O images (Figs. \ref{fig:intensity} and \ref{fig:intensity2}) that was not seen in $^{12}$CO \citep{vanderMarel2015-12co}. IRS48 shows a full gas ring in $^{13}$CO, which does not suffer from foreground absorption found in the $^{12}$CO \citep{Bruderer2014}. On the other hand, the foreground absorption seen in the $^{12}$CO emission of SR~21 \citep{vanderMarel2015-12co} also affects the $^{13}$CO spectra. 

\begin{figure}[!ht]
\subfigure{\includegraphics[scale=0.35,trim=0 0 50 0]{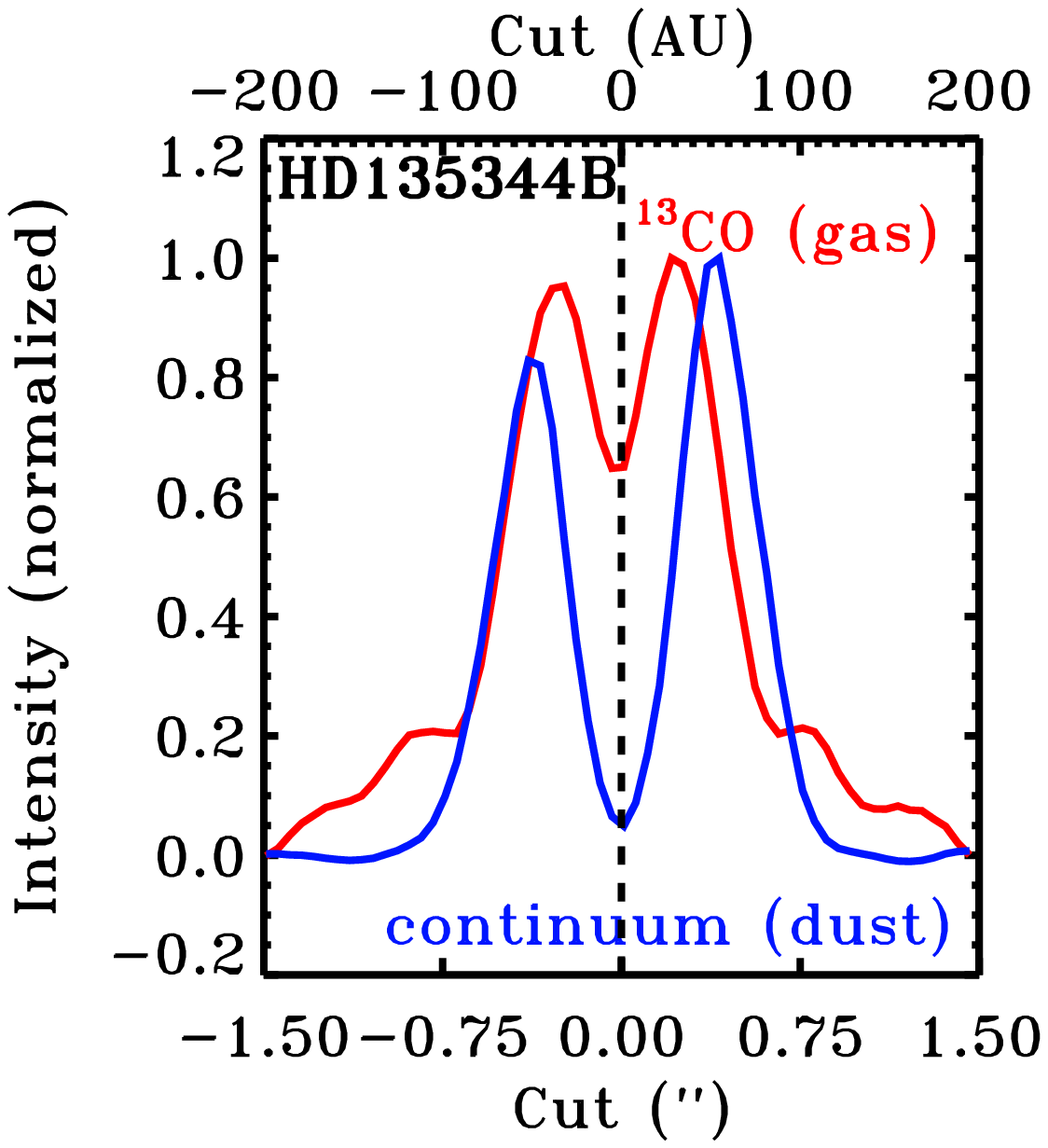}}
\subfigure{\includegraphics[scale=0.35,trim=50 0 0 0]{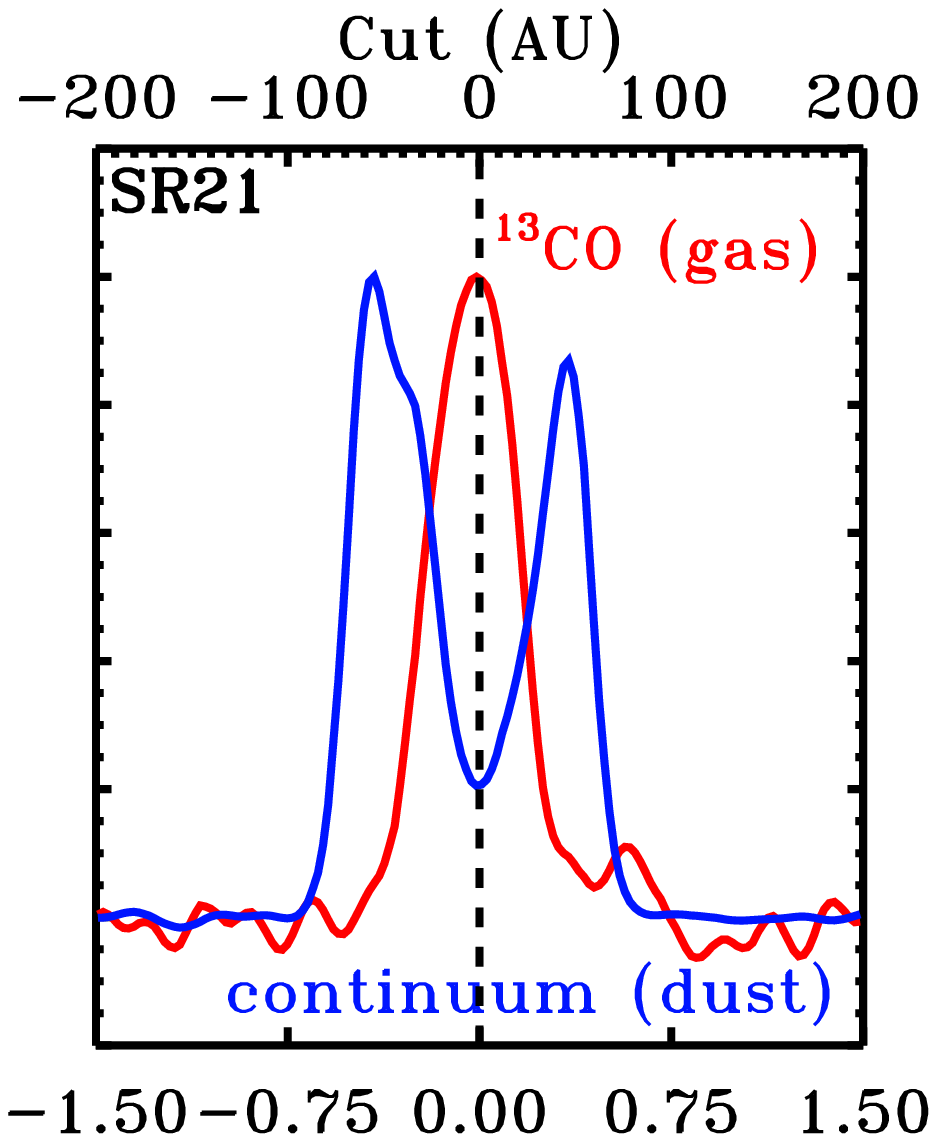}}\\
\subfigure{\includegraphics[scale=0.35,trim=0 0 50 0]{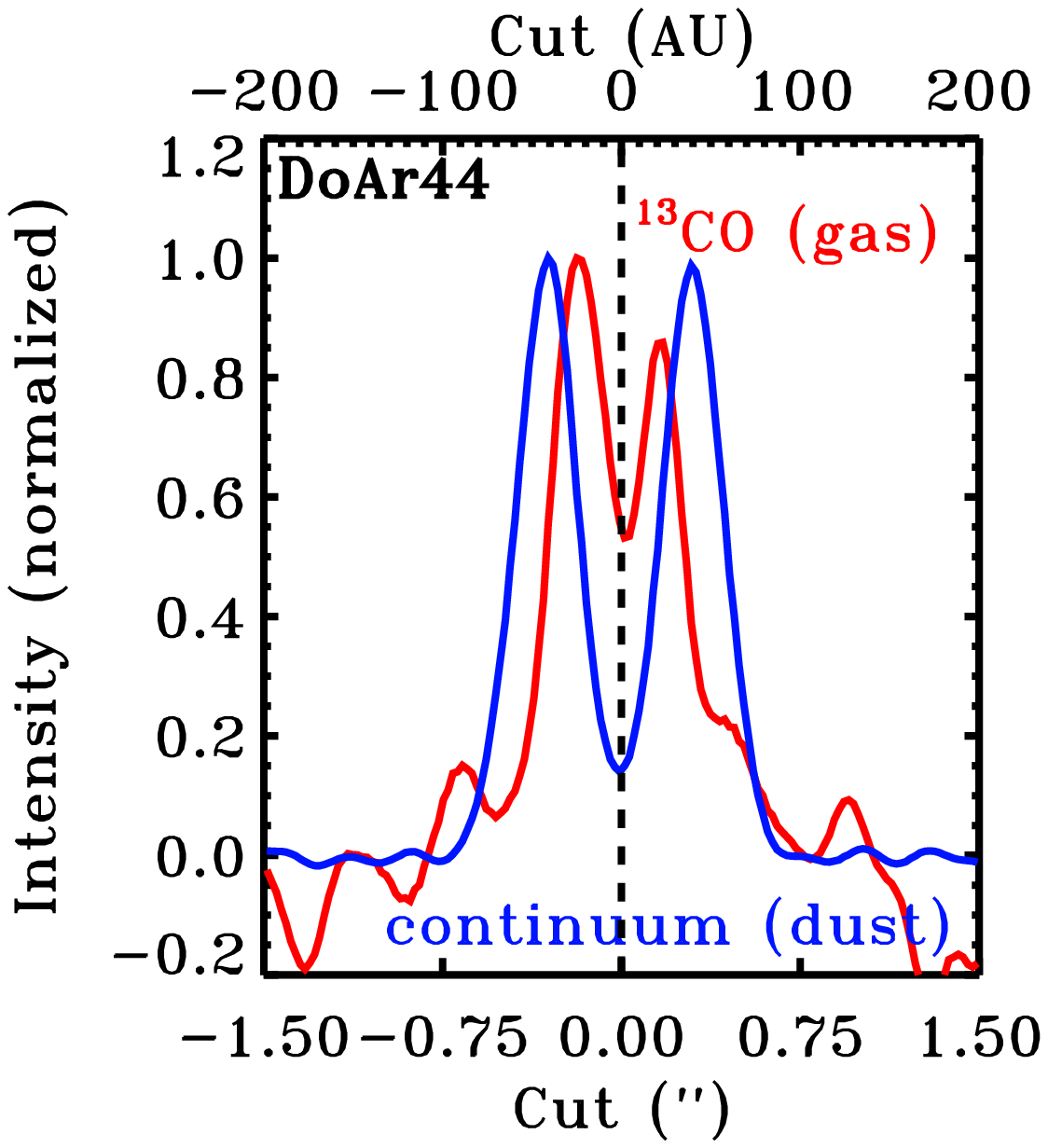}}
\subfigure{\includegraphics[scale=0.35,trim=50 0 0 0]{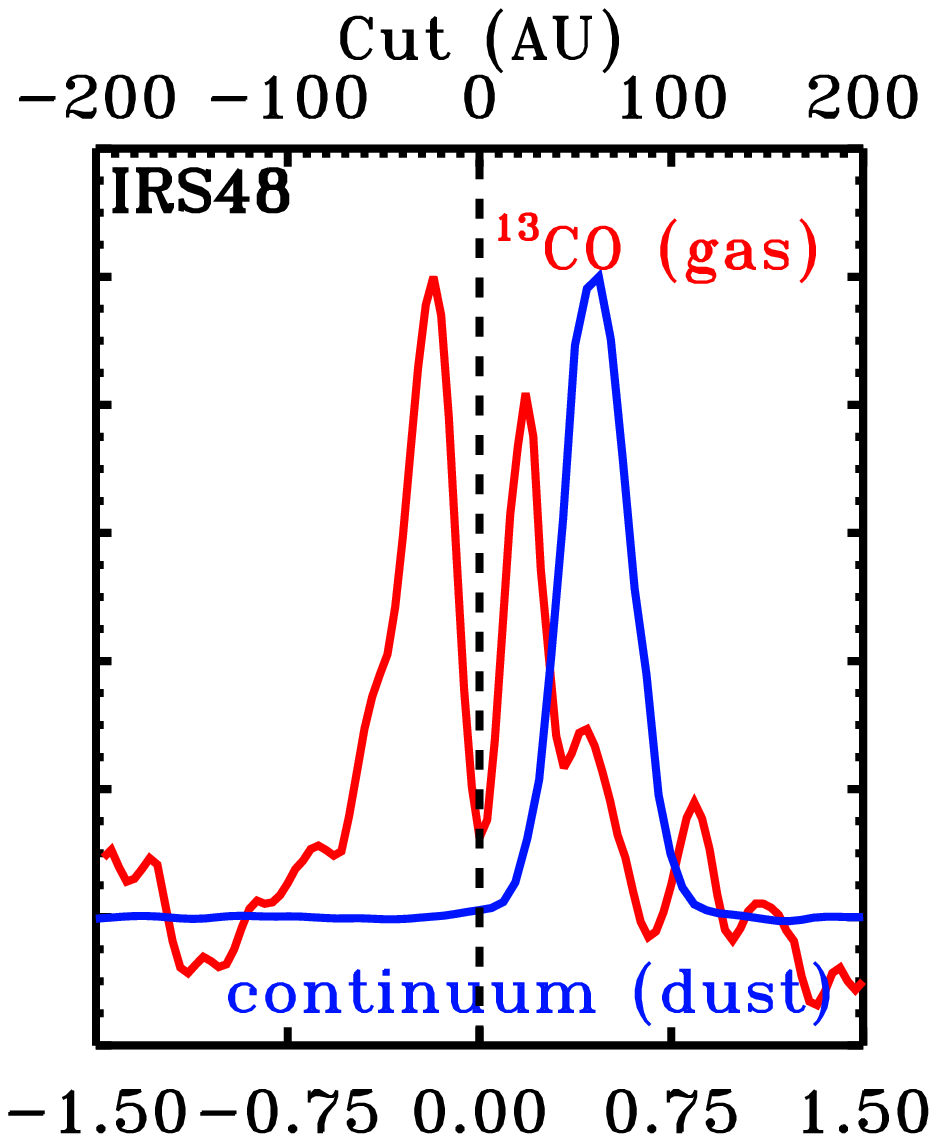}}
\caption{Normalized intensity cuts through the major axis of each disk of the $^{13}$CO 3-2 emission (red) and the dust continuum emission (blue). In case of IRS48, the deprojected intensity cut of the minor axis is taken so as to cover the (asymmetric) continuum profile. The cuts clearly reveal that the gas cavity radii are smaller than the dust cavity radii.}
\label{fig:intensitycuts}
\end{figure}

All CO data reveal rotating gas disks with inclination $>$15$^{\circ}$, with a double-peaked velocity profile. The gas rings for HD135344B, DoAr44, and IRS~48 are in all cases smaller than the continuum cavities, as shown directly in the $^{13}$CO panels in Figs. \ref{fig:intensity} and \ref{fig:intensity2} and in the radial cuts in Fig. \ref{fig:intensitycuts}. The emission inside the gas cavities is at least a factor of 2 lower than in the surrounding rings. 
In contrast, SR~21 does not appear to have a cavity in the gas at this spatial resolution. The peak S/N in the integrated maps is 12-30 for  $^{13}$CO and 5-20 for C$^{18}$O. 

The $^{13}$CO channel maps and velocity maps are used to derive the stellar position, the position angle, inclination, and source velocity, which are within errors consistent with the values derived from the $^{12}$CO data \citep{vanderMarel2015-12co}. The derived parameters are given in Table \ref{tbl:obsproperties}. 

The continuum images show ring-like structures for SR21, HD135344B, and DoAr44 and a highly asymmetric structure for IRS~48, as seen in previous Cycle 0 data \citep{vanderMarel2013}. The S/N on the continuum ranges between 130 and 500 for the different disks. The Band 7 continuum data of SR21 and HD135344B show minor azimuthal asymmetries with a contrast lower than a factor 2, similar to the Band 9 continuum \citep{Perez2014,vanderMarel2015-12co}, although the asymmetry in SR21 appears to be less pronounced in Band 7 than in Band 9. The Band 9 and Band 7 continuum images are compared and analyzed in \citet{Pinilla2015beta}. The continuum emission in IRS~48 is discussed in \citet{vanderMarel2015vla}. 

\section{Modeling}
\subsection{Physical model}
As a starting point for our models we adopted the physical structure suggested by \citet{Andrews2011}, as implemented by \citet{Bruderer2013} and fully described in \citet{vanderMarel2015-12co}. The surface density $\Sigma(r)$ is assumed to be a radial power-law with an exponential cut-off following the time-dependent viscosity disk model $\nu \sim r^{\gamma}$ with $\gamma=1$  \citep{LyndenPringle1974,Hartmann1998}
\begin{equation}
\Sigma(r) = \Sigma_c \left(\frac{r}{r_c}\right)^{-\gamma} {\rm exp}\left(-\left(\frac{r}{r_c}\right)^{2-\gamma}\right)
.\end{equation}
The gas and dust follow the same density profile, but the gas-to-dust ratio is varied throughout the disk, as shown in Fig. \ref{fig:genericmodel}. Inside the cavity, the dust density is zero, except for the inner disk, which is set by $\delta_{\rm dust}$. The gas density inside the cavity is varied with drops $\delta_{\rm gas}$. In the outer disk, the gas-to-dust ratio is fit by a constant number. The vertical structure is defined by the scale height $h_c$ and the flaring angle $\psi$, following $h(r)=h_c(r/r_c)^{\psi}$.  The fraction of large grains $f_{\rm ls}$ and the scale height of the large grains $\chi$ are used to describe the settling. More details on the star, the adopted stellar UV radiation, the dust composition, and vertical structure are given in \citet{vanderMarel2015-12co}.

\subsection{Model-fitting approach}
The best-fit models from Table 4 in \citet{vanderMarel2015-12co} were used as initial model for the vertical structure and dust density structure for SR21 and HD135344B, based on a combination of SED, dust 690 GHz continuum visibility, and $^{12}$CO 6--5 modeling. These models were fit by eye, starting from a surface density and cavity size consistent with the millimeter visibility curve, followed by small adjustments on the inner disk parameter ($\delta_{\rm dust}$) and vertical structure to fit the SED. For the fit to the $^{12}$CO data, the gas surface density was taken initially  assuming a gas-to-dust ratio of 100, and the amount of gas inside the cavity was subsequently constrained by varying the $\delta_{\rm gas}$ parameter, where $\Sigma_{\rm gas} = \delta_{\rm gas}\Sigma_{\rm gas}$ for $r<r_{\rm cav}$. The dust density inside the cavity (between $r_{\rm gap}$ and $r_{\rm cavdust}$) was set to be entirely empty of dust grains. SR21 is an exception: a small amount of dust was included between 7 and 25 AU, following \citet{vanderMarel2015-12co}.  The dust structure of DoAr44 is analyzed in a similar way in Appendix \ref{sct:doar44model} through SED and dust 345 GHz continuum visiblity modeling. For IRS48, we used the model derived by \citet{Bruderer2014}, although we chose to use an exponential power-law density profile instead of a normal power-law, to be consistent with the other three disks in this study. 

With the new CO isotopolog data, we used the initial dust structure model and only varied a small number of parameters to fit the CO emission by eye by subsequent adjustments of the surface density, gas-to-dust ratio, outer radius, and the $\delta$ parameters to fit the amount of gas inside the cavity. These are shown to be the most relevant parameters by our previous modeling. We did not use a $\chi^2$ or Markov chain Monte Carlo (MCMC) method
because the computational time of the models is too long and the number of parameters too large. 
Formal uncertainties of model parameters, the uniqueness of the fit and correlation between parameters cannot be computed directly, but the density and $\delta$ parameters are estimated to be within an order of magnitude and the radial parameters to within 5 AU based on a small model grid (see also Figs. \ref{fig:resulthd13} and \ref{fig:changecavitysize}).

\begin{figure}[t]
\includegraphics[scale=0.3]{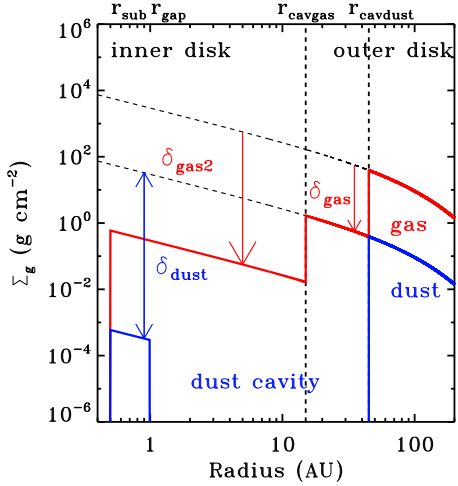}
\caption{Generic surface density profile for the gas and dust.}
\label{fig:genericmodel}
\end{figure}

The outer disk CO emission and submillimeter continuum flux were fit simultaneously by varying $\Sigma_c$ and the gas-to-dust
ratio. The outer radius $r_{\rm out}$ was set by fitting the CO spectrum (the maxima in the spectra or double peaks that arise from the Keplerian motion) and provides outer boundaries for computing the gas masses. Using this surface density profile, the emission inside the cavity is constrained by the $\delta$ parameters (Fig. \ref{fig:genericmodel}). The near-infrared excess determines the dust density in the inner disk through $\delta_{\rm dust}$. The gas cavity radius $r_{\rm cavgas}$ and drop $\delta_{\rm gas2}$ were fit to the CO emission. In some cases, an additional drop in gas surface density is required between $r_{\rm cavgas}$ and the dust cavity radius $r_{\rm cavdust}$. This drop is indicated by $\delta_{\rm gas}$.

\begin{table}\small
\caption{Results for the gas density profile of each transition disk.}
\label{tbl:physmodel}
\begin{tabular}{llllll}
\hline
\hline
&Parameter&HD&SR&DoAr&IRS\\
&&13$^{(a)}$&21&44&48\\
\hline
Surface&$r_c$ (AU)&25&15&25&60\\
density&$\Sigma_c$ (g cm$^{-2}$)&120&400&60&0.5\\
&gas-to-dust ratio&80&100&100&12\\
&$M_{\rm dust}^{(b)}$ (10$^{-3} M_{\odot}$)&0.13&0.075&0.05&0.015\\
&$M_{\rm gas}^{(b)}$ (10$^{-3} M_{\odot}$)&15&7.7&2.5&0.55\\
\hline
Radial&$r_{\rm cavgas}$ (AU) $^{(d)}$&30&7&16&25\\
structure&$\delta_{\rm gas2}$ $^{(d)}$&2$\cdot10^{-4}$&$10^{-20}$ $^{(c)}$&$\leq10^{-4}$&$\leq10^{-3}$\\
&$r_{\rm cavdust}$ (AU)&40&25&32&60\\
&$\delta_{\rm gas}$ $^{(d)}$&1&$5\cdot10^{-2}$&$10^{-2}$&1\\
&$r_{\rm out}$ (AU)&125&60&60&90\\
\hline
Inner&$r_{\rm sub}$ (AU)&0.18&0.18&0.08&0.4\\
disk&$r_{\rm gap}$ (AU)&0.25&1.0&1.0&1.0\\
&$\delta_{\rm dust}$ (AU)&2$\cdot10^{-4}$&1$\cdot10^{-6}$&1$\cdot10^{-2}$&1$\cdot10^{-3}$\\
\hline
Vertical&$h_c$ (rad)&0.15&0.07&0.1&0.14\\
structure&$\psi$ &0.05&0.15&0.1&0.22\\
&$f_{\rm ls}$ &0.95&0.85&0.85&0.85\\
&$\chi$ &0.8&0.2&0.2&0.2\\
\hline
\end{tabular}
\tablefoot{
a) HD13=HD135344B.
b) The masses are only constrained within $r_{\rm out}$ for the detected surface brightness.
c) The drop inside 7 AU could not be constrained by the ALMA data; we adopt the value from CO rovibrational spectroastrometry by \citet{Pontoppidan2008}. 
d) The uncertainties on the gas cavity size are typically $\lesssim$ 5 AU and lower than one order of magnitude on the depth of the drop.
}
\end{table}

The main parameters that are varied are the gas cavity radius $r_{\rm cavgas}$ and the drop in gas density $\delta_{\rm gas2}$ between $r_{\rm sub}$ and $r_{\rm cavgas}$. The CO isotopologs provide better constraints on the density than the $^{12}$CO because they are less optically thick ($^{13}$CO) or even optically thin (C$^{18}$O). The effects of isotope-selective photodissociation are properly considered in the modeling and are discussed in Sect. \ref{sct:photodiss}. 

\section{Results}
\begin{figure*}
\begin{center}
\subfigure{\includegraphics[scale=0.4,trim=0 0 0 0]{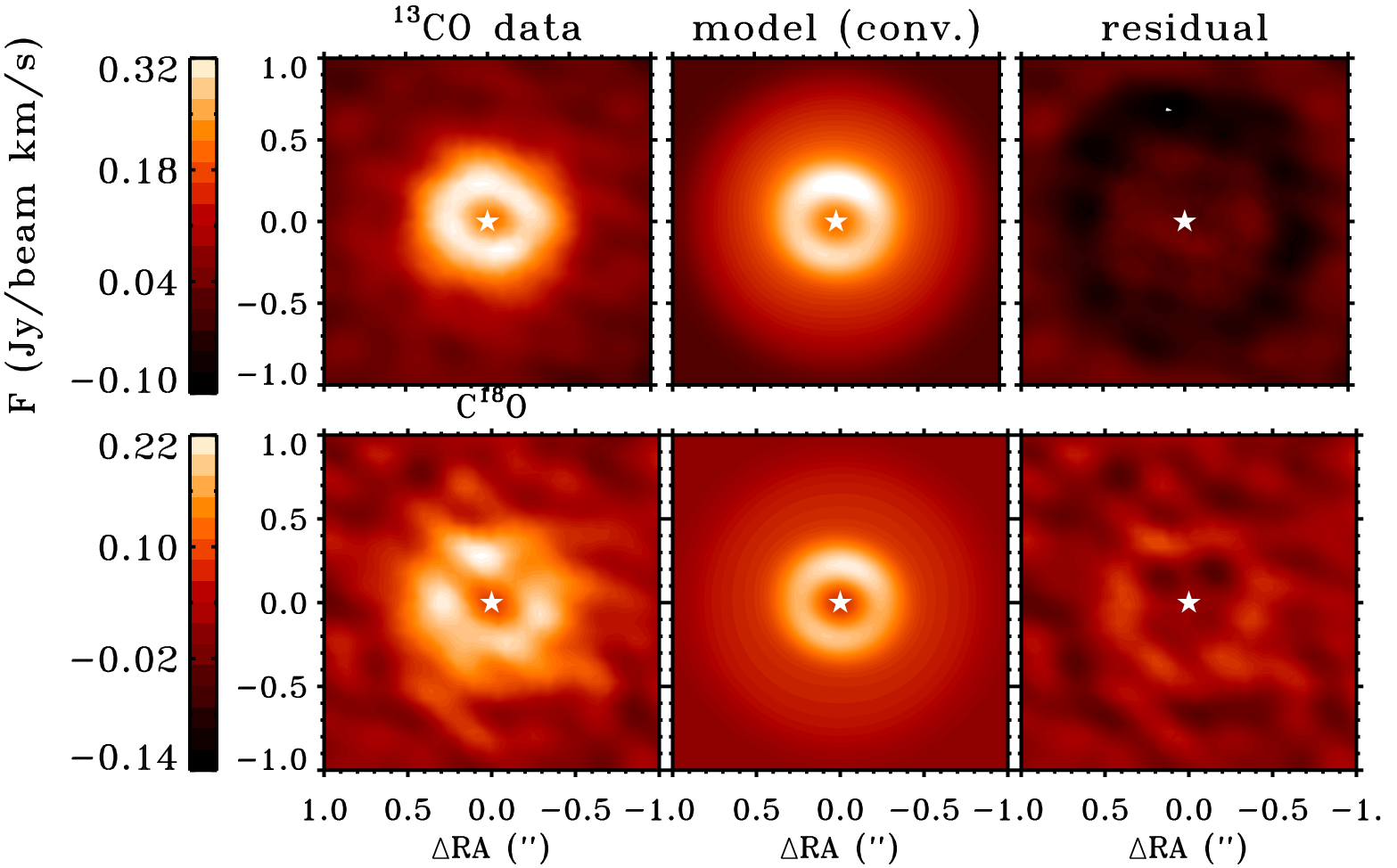}}
\subfigure{\includegraphics[scale=0.52,trim=28 40 20 40]{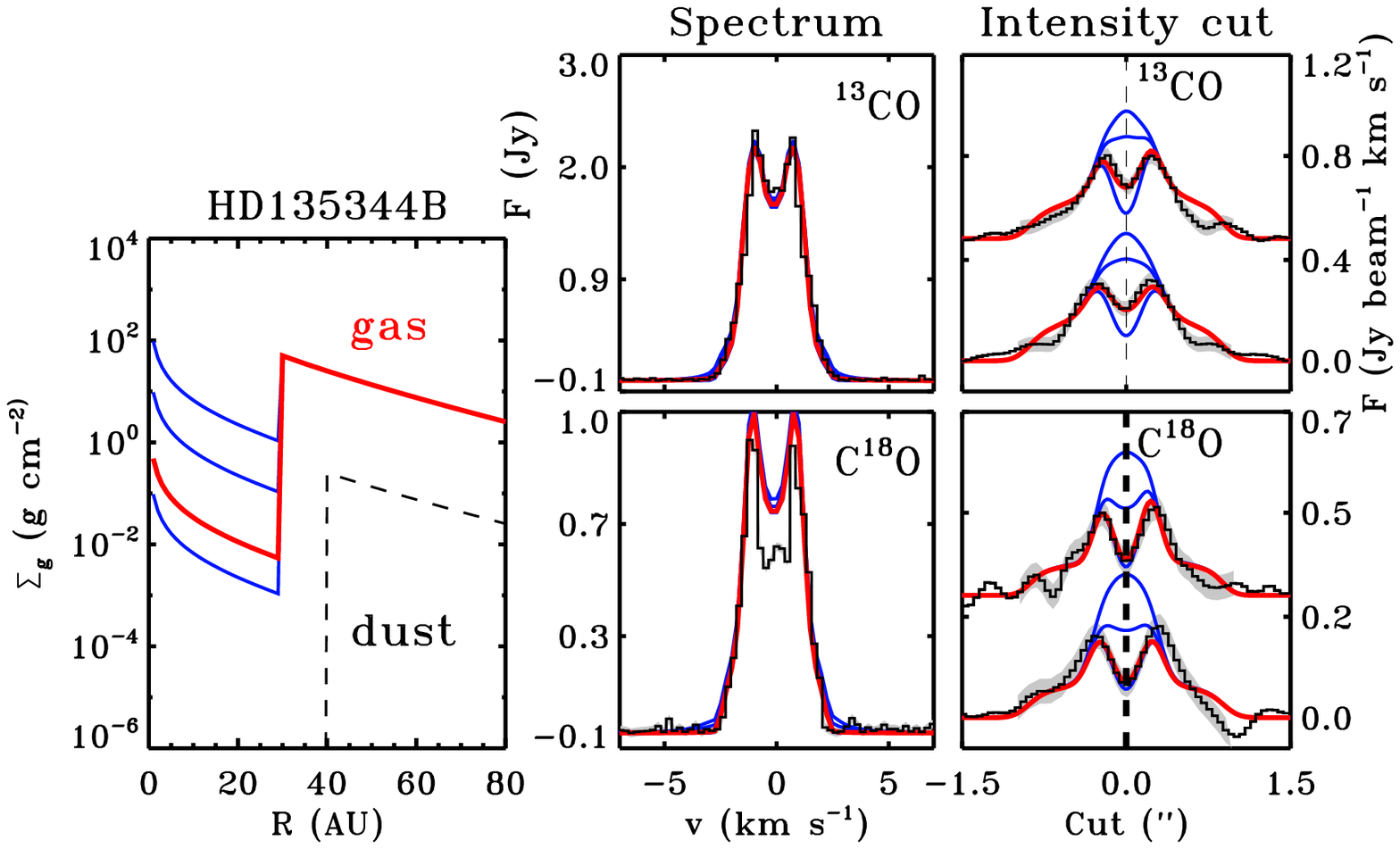}}

\subfigure{\includegraphics[scale=0.4,trim=10 20 10 0]{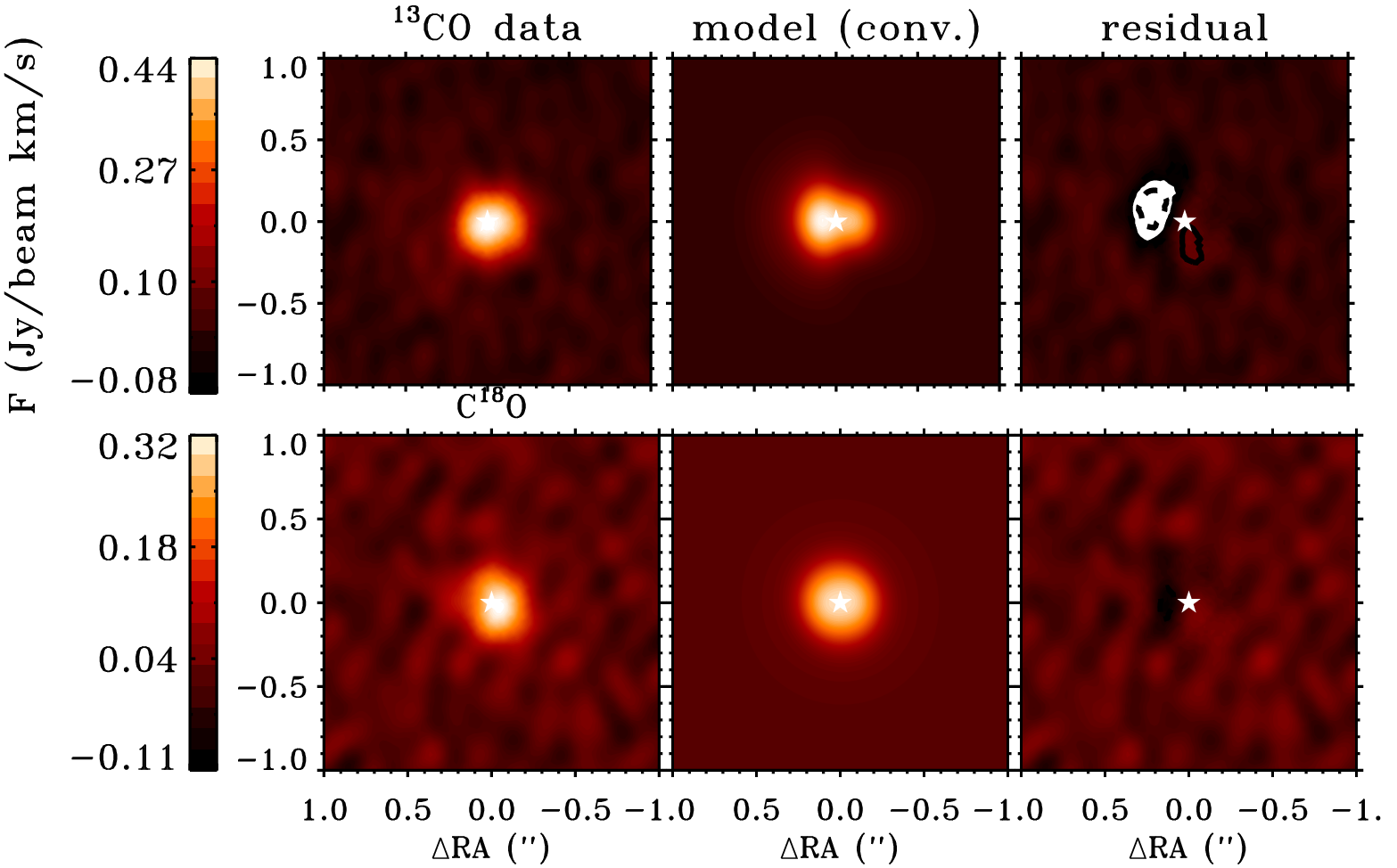}}
\subfigure{\includegraphics[scale=0.52,trim=20 40 20 40]{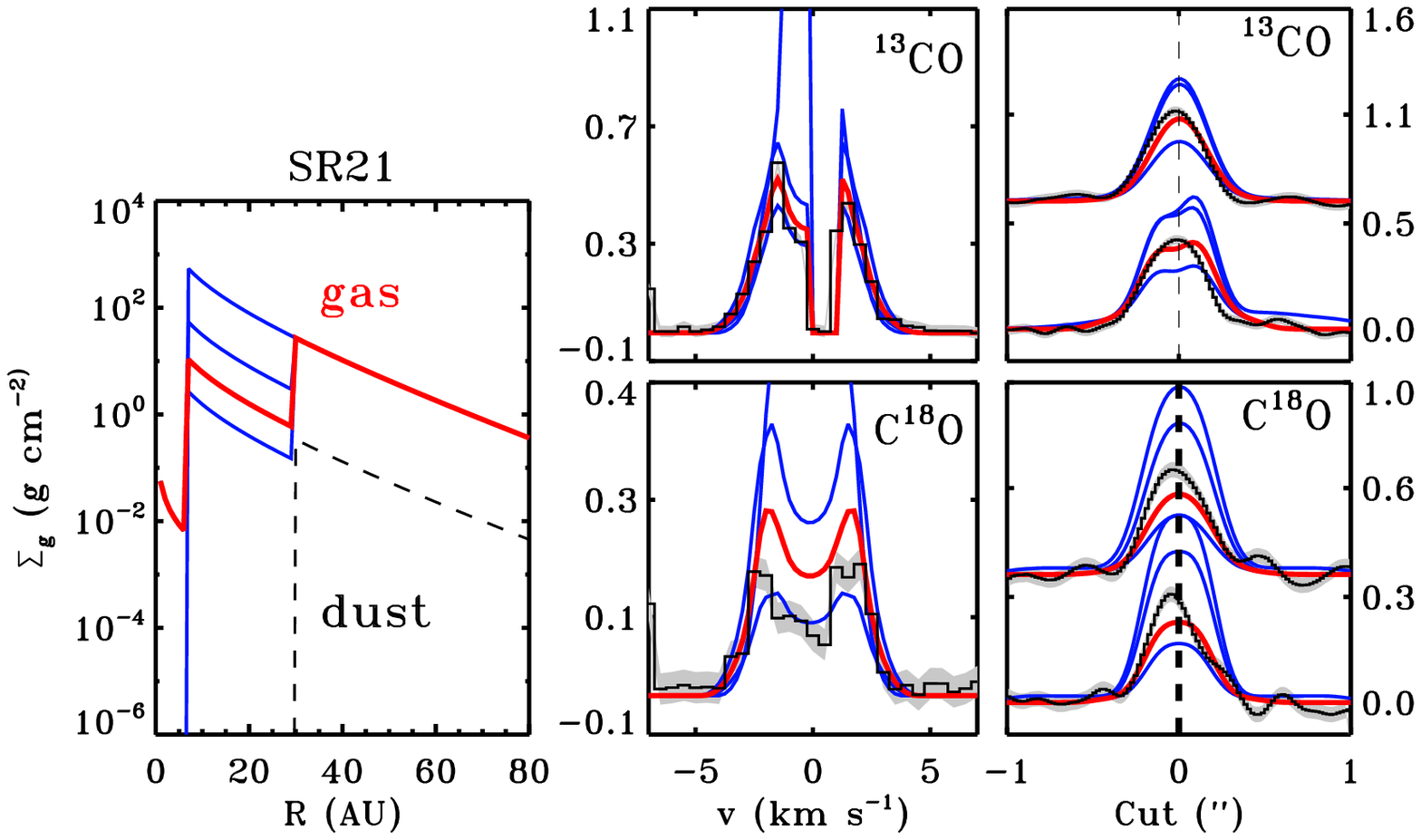}}

\subfigure{\includegraphics[scale=0.4,trim=10 20 10 0]{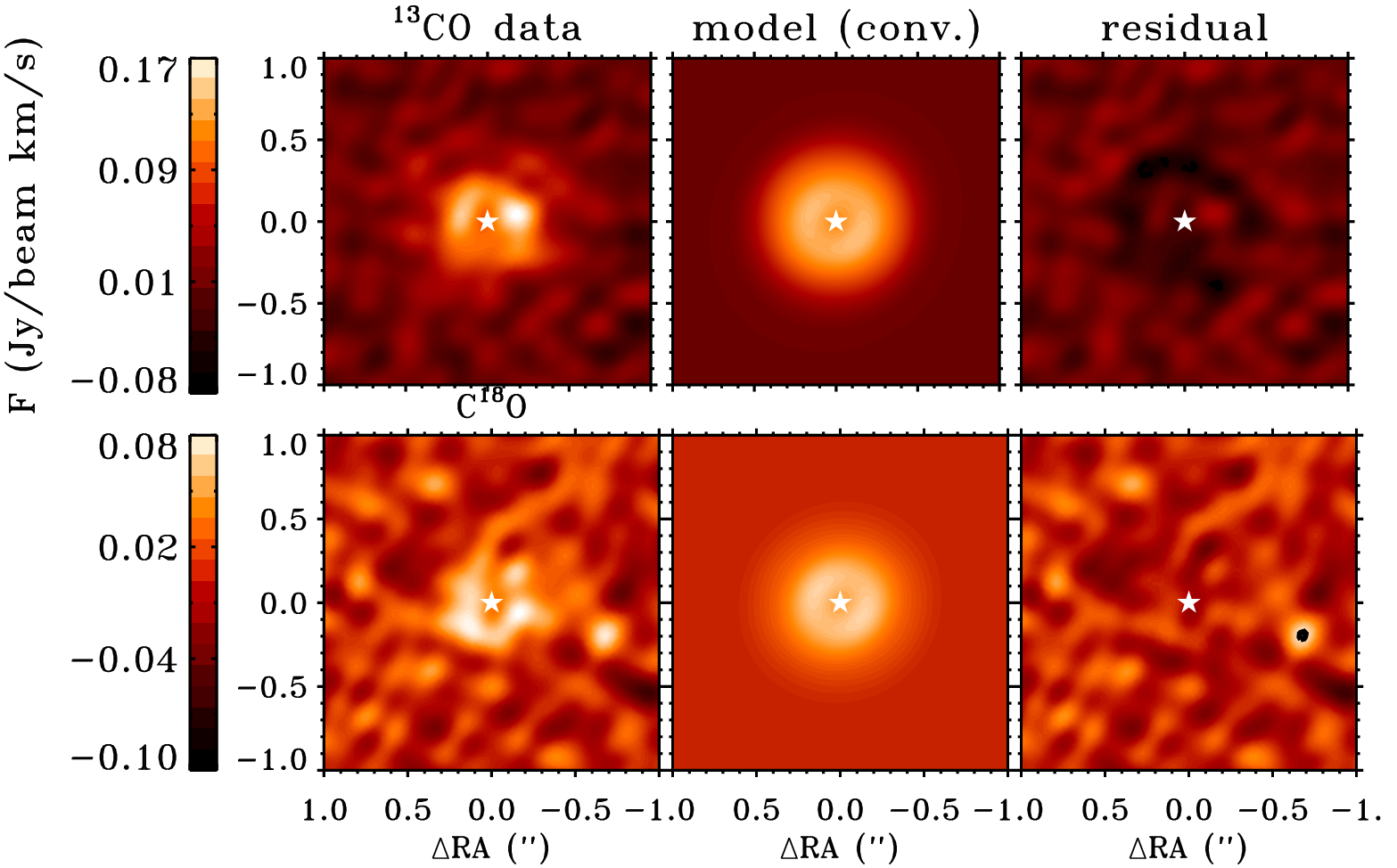}}
\subfigure{\includegraphics[scale=0.52,trim=20 40 20 40]{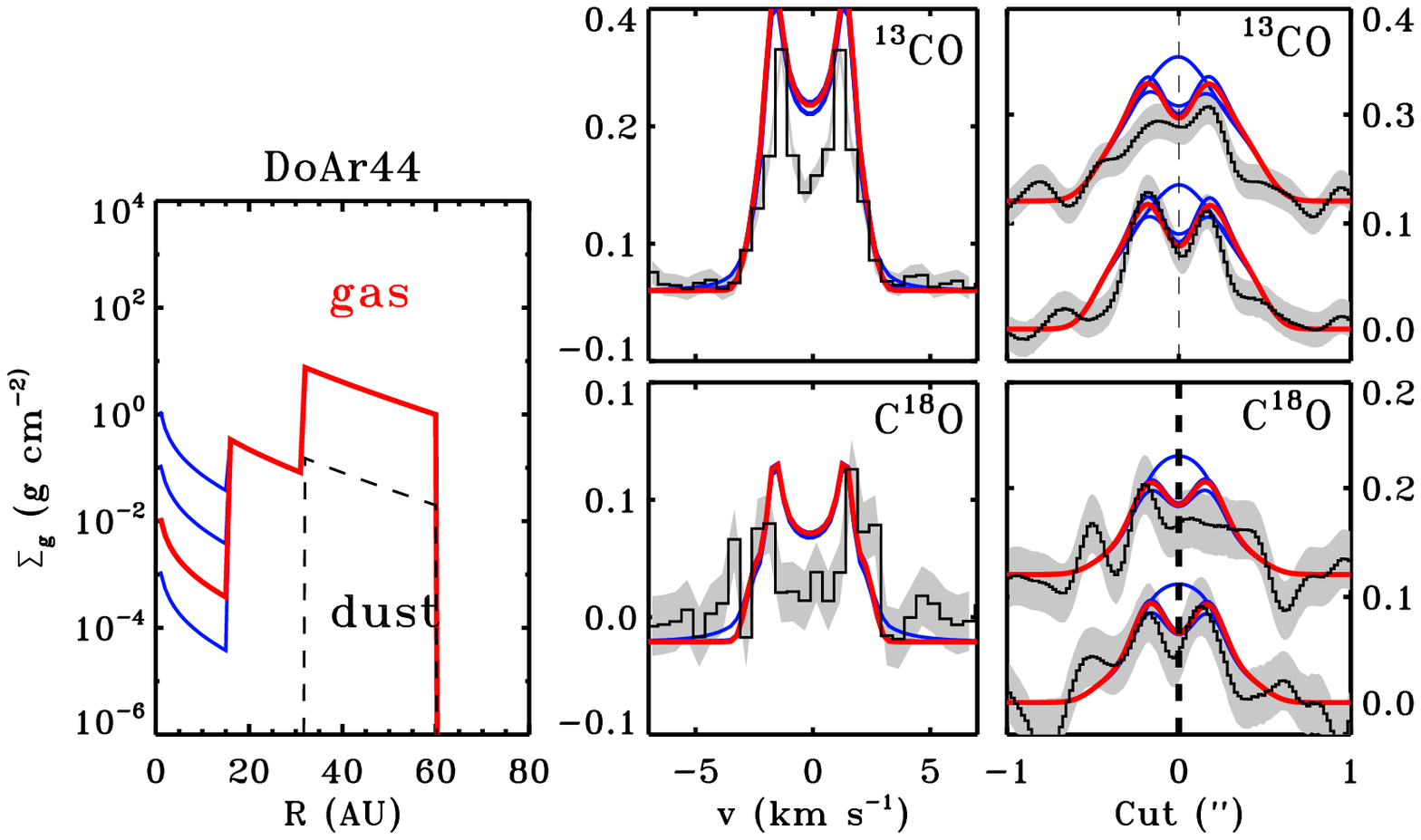}}

\subfigure{\includegraphics[scale=0.4,trim=10 20 0 0]{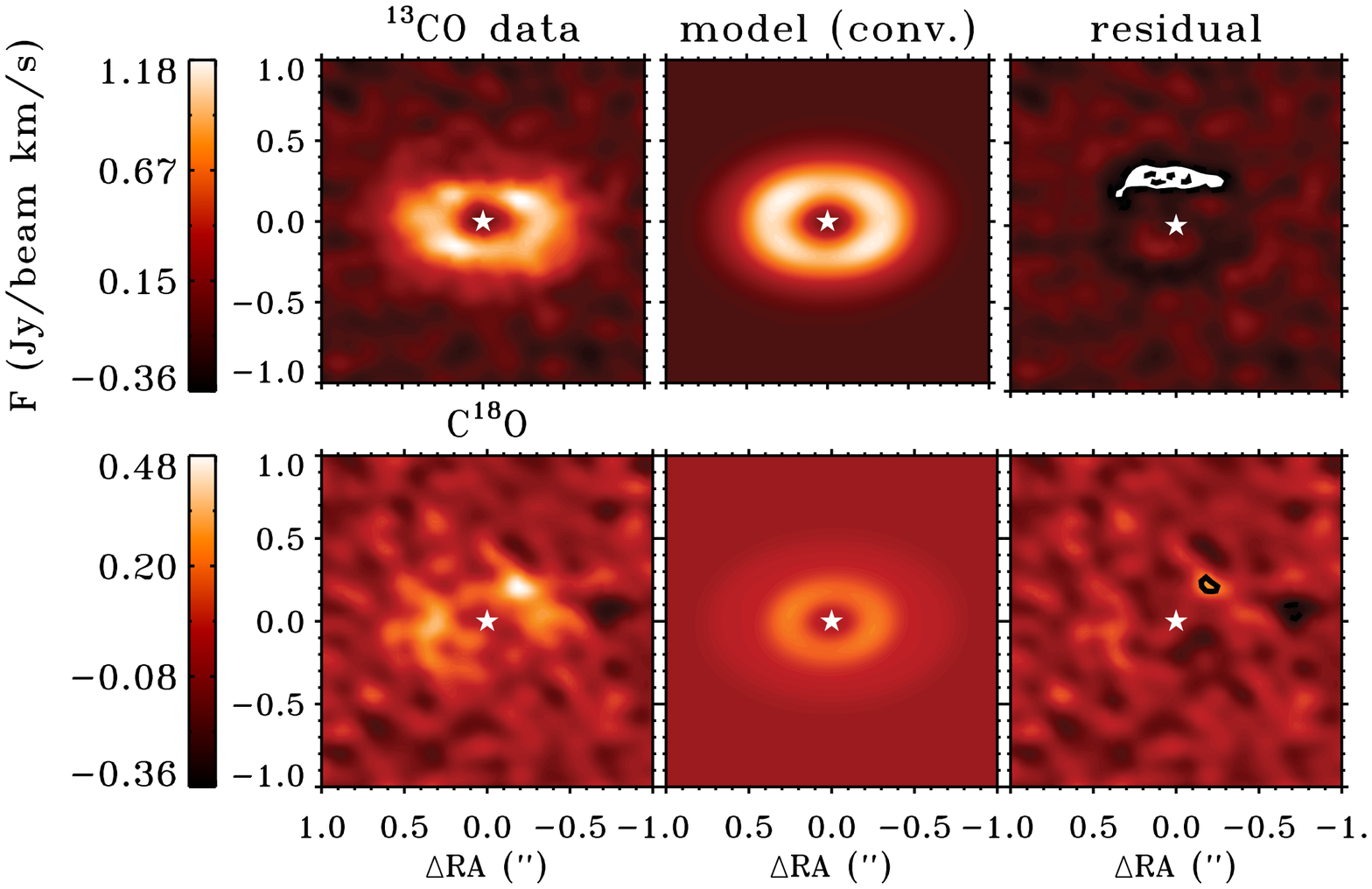}}
\subfigure{\includegraphics[scale=0.52,trim=20 40 20 40]{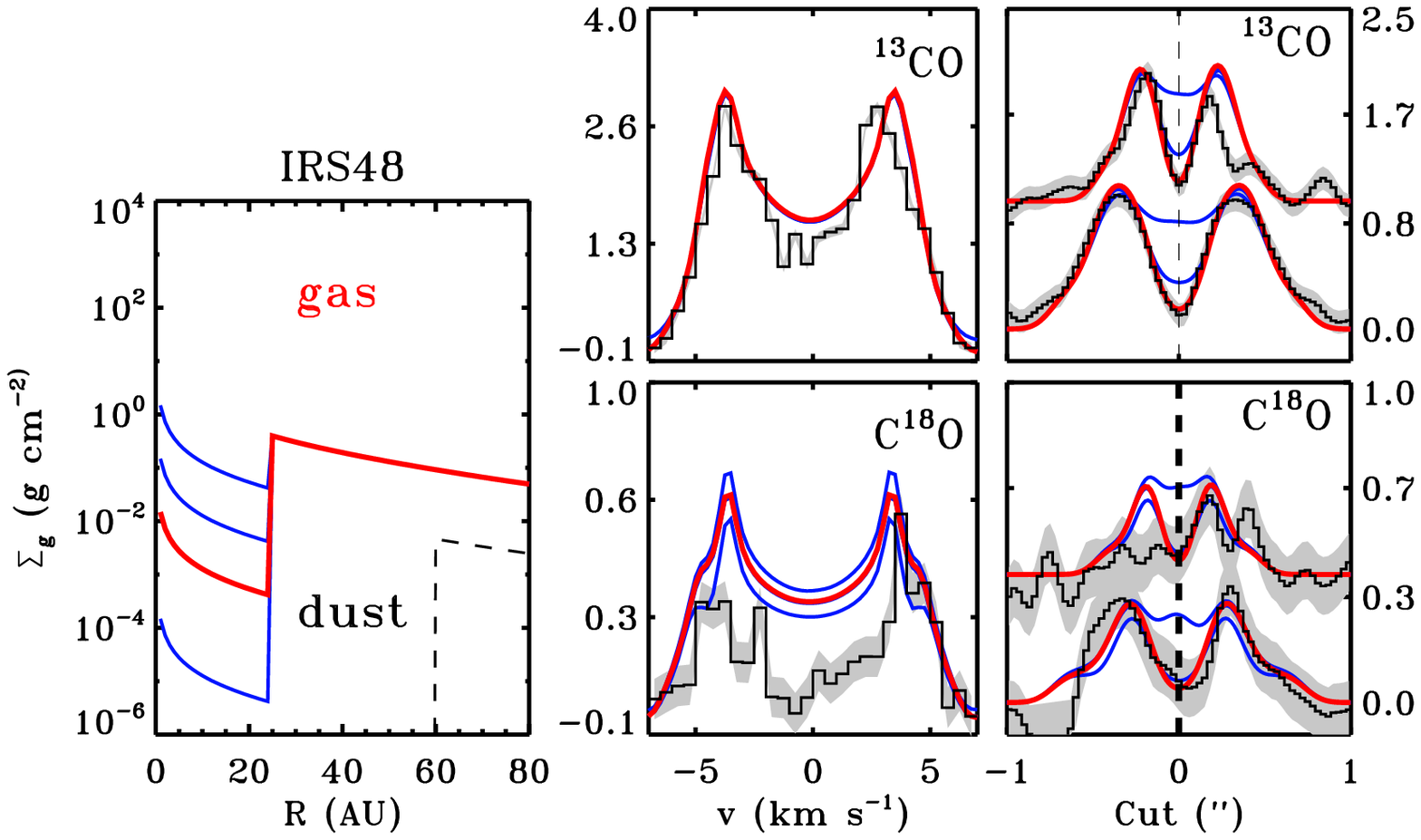}}
\end{center}
\caption{Modeling results and observations of the $^{13}$CO and C$^{18}$O emission for the best-fitting models in derotated images and spectra. The left panels show the direct image comparison of the zero-moment map: derotated observations, model, convolved model, and residual. In the residual map the overlaid contours indicate the 3$\sigma$ levels, where dashed lines are negative. The central panels show the gas density profiles with different $\delta_{\rm gas2}$ drops as solid lines and the dust density profile as dashed lines. The right panels show the resulting intensity cuts for the major and minor axis for both $^{13}$CO and C$^{18}$O. The best-fitting model is indicated in red, other $\delta_{\rm gas2}$ values in blue. The drop in density can be constrained to within an order of magnitude.}
\label{fig:resulthd13}
\end{figure*}
Data and models are compared through the spectra and the zero-moment maps (intensity maps) of both $^{13}$CO and C$^{18}$O. For each target we show the direct comparison of the images of the intensity map of the best-fit model and the constraint on both $r_{\rm cavgas}$ and $\delta_{\rm gas2}$ through spectra and intensity cuts through the major and minor axis of the zero-moment map.

In three of the four targets an inner gas cavity (smaller than the dust cavity) was required to fit the data. In SR21 the 7 AU cavity previously derived from rovibrational emission was adopted, but no gas cavity is resolved at the spatial resolution of $\sim0.2"$ (24 AU) in our observations. In all disks, the depth is constrained to within an order of magnitude and the cavity size to within $\pm$5 AU, mainly through the comparison of the spatially resolved emission, but further confirmed by the line wings in the spectra. In the intensity cuts in Fig. \ref{fig:resulthd13} the $\delta_{\rm gas2}$ parameter is varied. The comparison shows that the uncertainty on the density drop is lower than an order of magnitude. In Appendix \ref{sct:additionalmodels} we show the intensity cuts for larger and smaller gas cavity radii, respectively. These plots show that the uncertainty on the gas cavity radius is typically $\lesssim$5 AU and the uncertainty on the depth is lower than one order of magnitude. Furthermore, in both SR21 and DoAr44 an additional drop $\delta_{\rm gas}$ between $r_{\rm cavgas}$ and $r_{\rm cavdust}$ is required to fit the data. 

The models of the $^{12}$CO fits of \citet{vanderMarel2015-12co} match the isotopolog data in the outer disk remarkably well. Only minor corrections in surface density and gas-to-dust ratio were required to fit the new data. However, inside the cavity the previously derived results for the gas surface density were found to be inconsistent. A modest drop in the gas density inside the dust cavity radius $r_{\rm cav}$ was sufficient to explain the $^{12}$CO data, but no gas cavity was seen. Since the gas cavity becomes visible in the isotopolog data and turns out to be smaller than the dust cavity radius, the gas cavity radius $r_{\rm cavgas}$ could be fit independently of the depth.

\subsection{Results for individual targets}
\paragraph{HD135344B\\}
HD135344B shows a clear gas cavity, both in $^{13}$CO and C$^{18}$O images, which is significantly smaller than the dust cavity (see top two panels of Fig. \ref{fig:intensity}). The modeling constrains the gas cavity radius to 30 AU (dust cavity radius is 40 AU), and the drop is $2\cdot10^{-4}$ (top right panels in Fig. \ref{fig:resulthd13}). Previously, the model based on the $^{12}$CO data required $\delta_{\rm gas}$ of 10$^{-1}-10^{-2}$ for a cavity radius of 40 AU. A 30 AU gas cavity was also derived in \citet{vanderMarel2015-12co} considering the inner radius of the small dust grains derived from the scattered-light emission \citep{Garufi2013}, but a $\delta_{\rm gas}$ value as low as $\sim10^{-4}$ underproduces the $^{12}$CO emission inside the cavity. As the $^{12}$CO is optically thick and mainly traces the temperature at the $\tau=1$ surface, this new result suggests that the temperature structure is somewhat higher than in our model, perhaps because of some residual dust in the cavity  (increasing the CO abundance by its shielding and changing the heating-cooling balance) or the vertical structure (shadowing on the outer disk) \citep[see][for a detailed discussion]{Bruderer2013}.

The residual of the $^{13}$CO emission (left panels in Fig. \ref{fig:resulthd13}) shows that the model overpredicts the emission in the outer disk ($>$0.5"). On the other hand, the C$^{18}$O residual underpredicts the emission in the outer disk. The residuals are spatially insignificant, but integrated over the whole disk, they are most likely related to real structure,  suggesting an outer gas ring with a possible gap. 
The radius of $\sim100$ AU coincides with a suggested planet that is launching one of the spiral arms observed in scattered-light observations \citep{Muto2012}. If there is indeed a planet at this large orbit, it is expected to clear a gap in the gas, which might explain the structure in the outer disk. As the focus of this study is on the emission inside the dust cavity, this is not investigated further. 

\paragraph{SR21\\}
Unlike the other disks, the CO emission in SR21 does not appear to have a cavity. Rovibrational CO emission suggests a gas cavity of 7 AU radius \citep{Pontoppidan2008} that remains unresolved at the ALMA spatial resolution (14 AU diameter$\sim$0.12''). The physical model includes this 7 AU cavity. The $^{12}$CO emission already indicated a drop in density between 7 AU and the dust cavity radius of 25 AU. The intensity cuts of the CO isotopologs also suggests a drop in density considering the strength of the emission (right panels in Fig. \ref{fig:resulthd13}). According to the model fitting, this drop is of about two orders of magnitudes. 

\paragraph{DoAr44\\}
DoAr44 is the only disk in this study with a symmetric dust ring. The dust cavity radius was found to be 32 AU (see Appendix \ref{sct:doar44model}), using the same modeling approach as in \citet{vanderMarel2015-12co}. The cavity size and disk mass are similar to previous studies of SMA 345 GHz continuum data \citep{Andrews2011}. The drop in dust density inside the cavity ($\delta_{\rm dustcav}$) is at least a factor 1000. The CO isotopolog intensity maps have a
lower S/N than the other disks, probably because of the lower disk mass, but still show a gas cavity that is only half the size of the dust cavity: 16 AU. The $\delta_{\rm gas2}$ inside 16 AU is at most 10$^{-4}$. An additional drop in gas density between 16 and 32 AU of 10$^{-2}$ is required to fit the emission (right panels in Fig. \ref{fig:resulthd13}). 

\paragraph{IRS48\\}
The CO isotopolog emission in IRS48 confirms the presence of a gas cavity with a considerably smaller radius than that of the peak of the dust asymmetry (60 AU), as found by \citet{Bruderer2014}, consistent with the 30 AU gas ring found in rovibration CO emission \citep{Brown2012a}. To be consistent with the models of the other disks, we assumed a radial power-law with exponential cut-off instead of a simple power-law as done by \citet{Bruderer2014}, but a similar vertical structure, density structure, and mass are found as in their study. Because DALI considers only axisymmetric models, the millimeter continuum asymmetry is not fit: the total submillimeter flux, the VISIR 18$\mu$m image and the SED are again used to constrain the dust density and dust cavity radius. The gas cavity radius is found to be 25 instead of the 20 AU found by \citet{Bruderer2014}, but this is within the uncertainty on the radius, and no additional drop between 25 and 60 AU in density is required to fit the emission (right panels in Fig. \ref{fig:resulthd13}). 

$^{13}$CO shows a full ring of emission. However, the emission is weaker at the location of the dust trap, as seen in the residual image (left panels in Fig. \ref{fig:resulthd13}). The $^{13}$CO emission is barely optically thick at the dust trap radius of 60 AU (the $\tau=1$ surface is at the midplane), and the drop in emission is most likely due to the continuum optical depth or a drop in temperature caused by the local increase of dust density. The S/N of the C$^{18}$O emission is too low to show the gas ring or this local drop in emission, but the data were used to set constraints on the gas density.

\subsection{Gradual drop}
\begin{figure}[!ht]
\subfigure{\includegraphics[scale=0.4,trim=0 55 0 0]{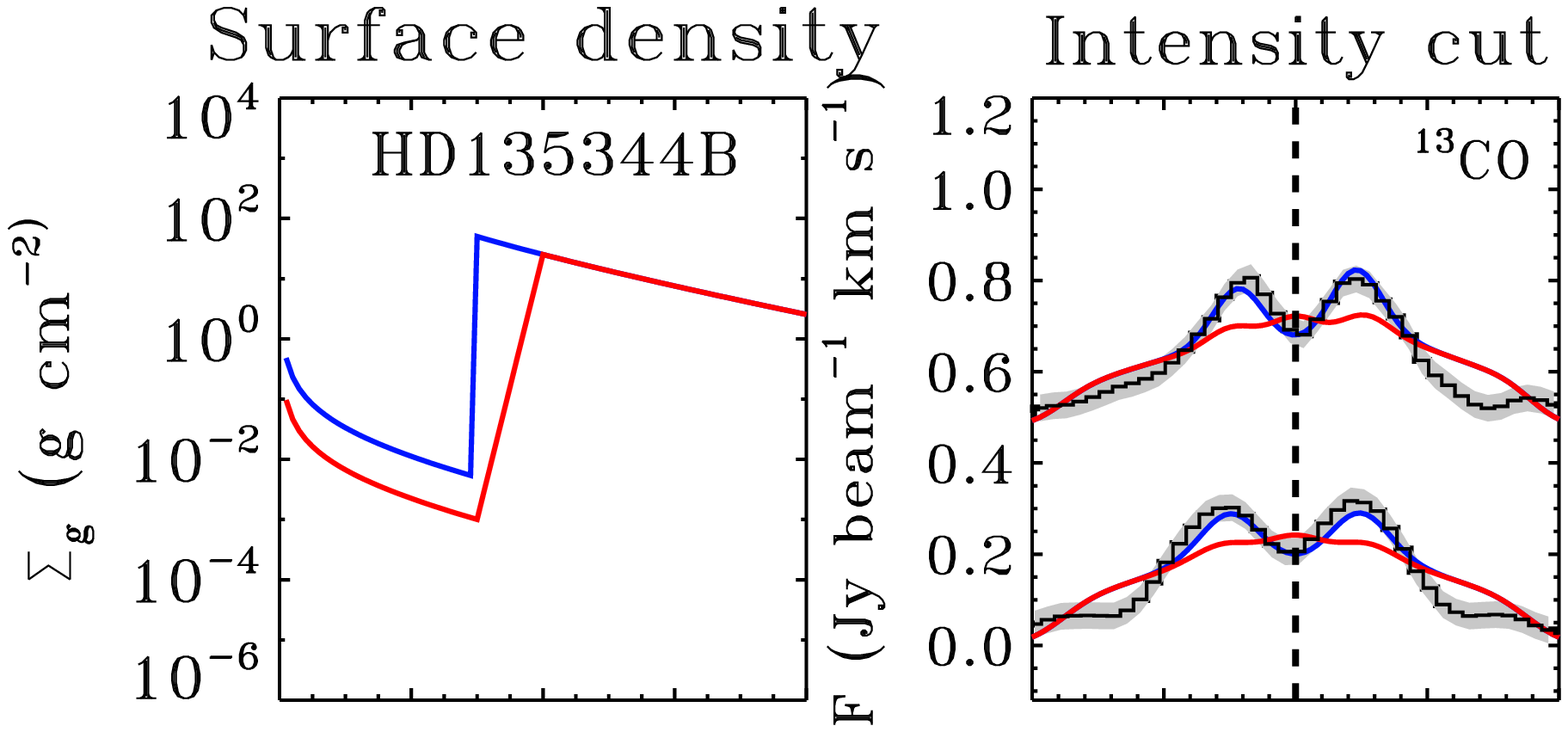}}
\subfigure{\includegraphics[scale=0.4,trim=0 55 0 20]{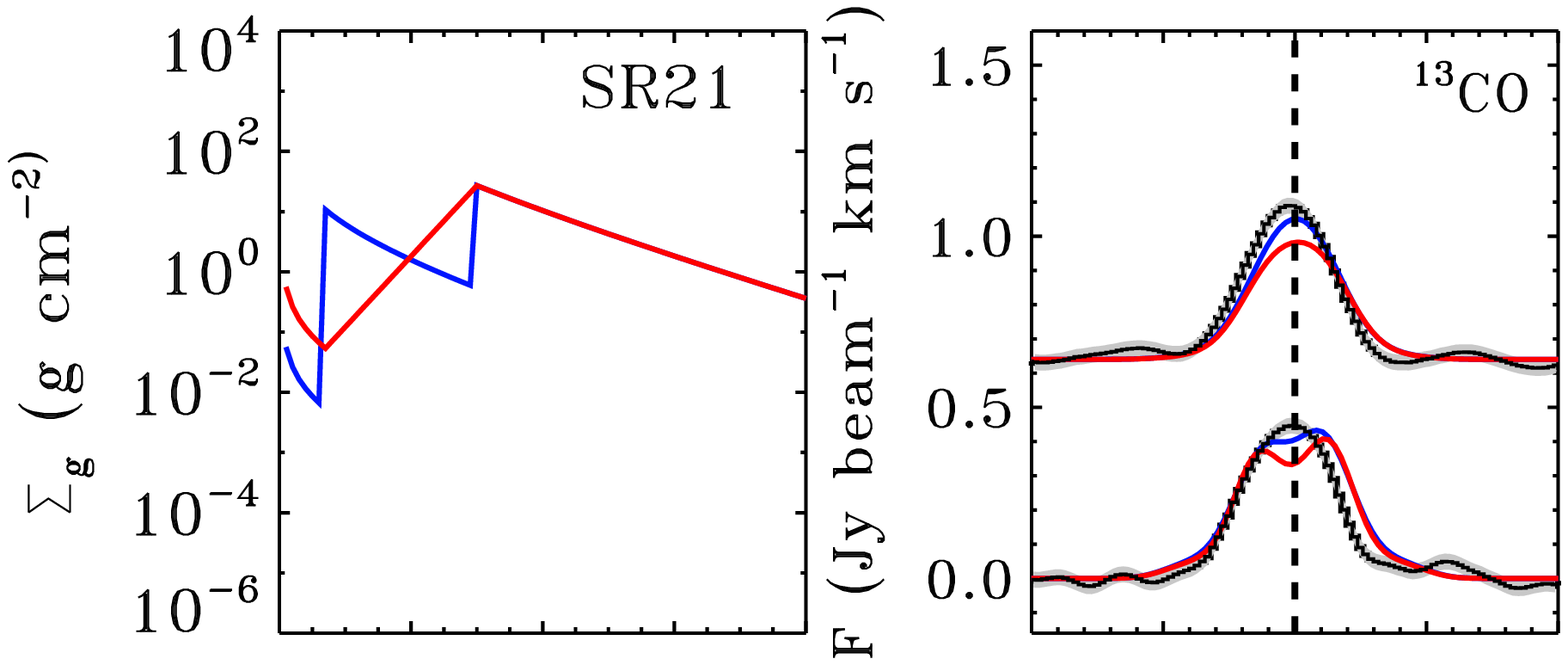}}
\subfigure{\includegraphics[scale=0.4,trim=0 55 0 20]{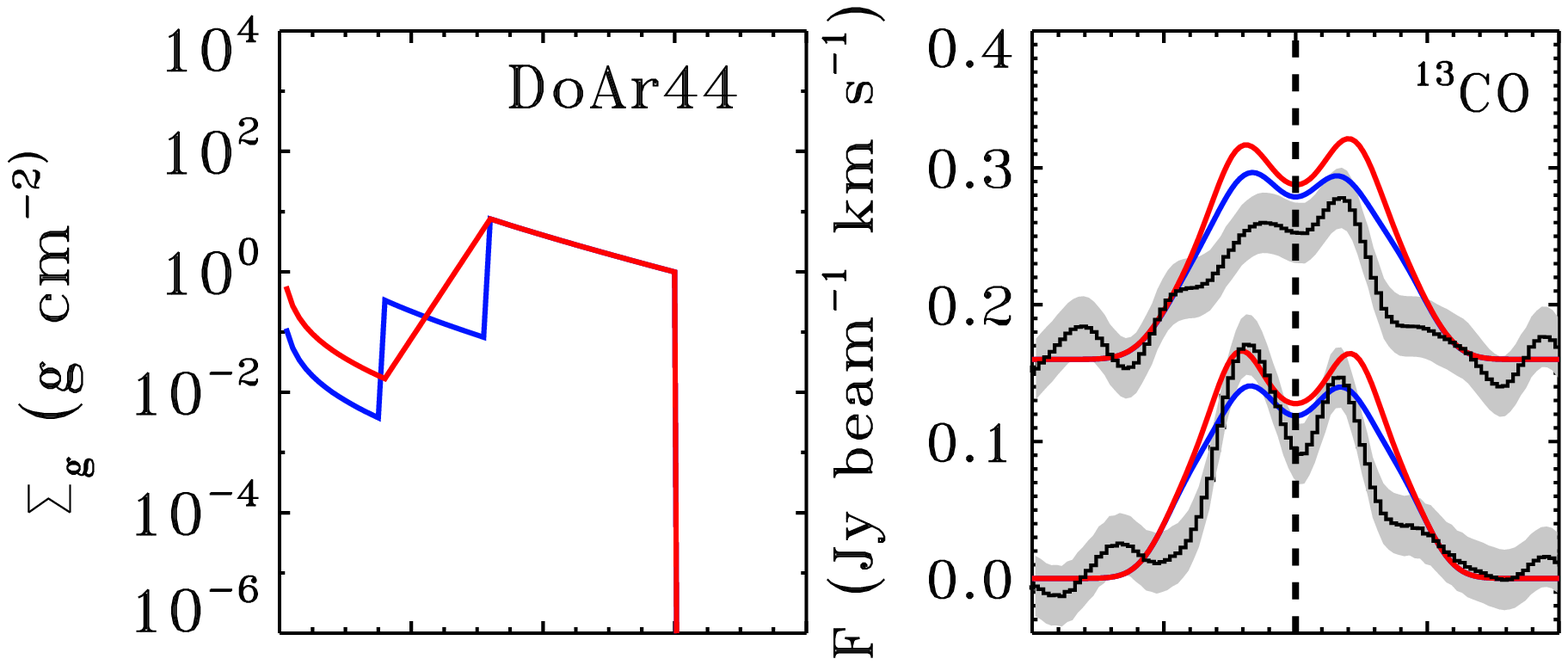}}
\subfigure{\includegraphics[scale=0.4,trim=0 0 0 20]{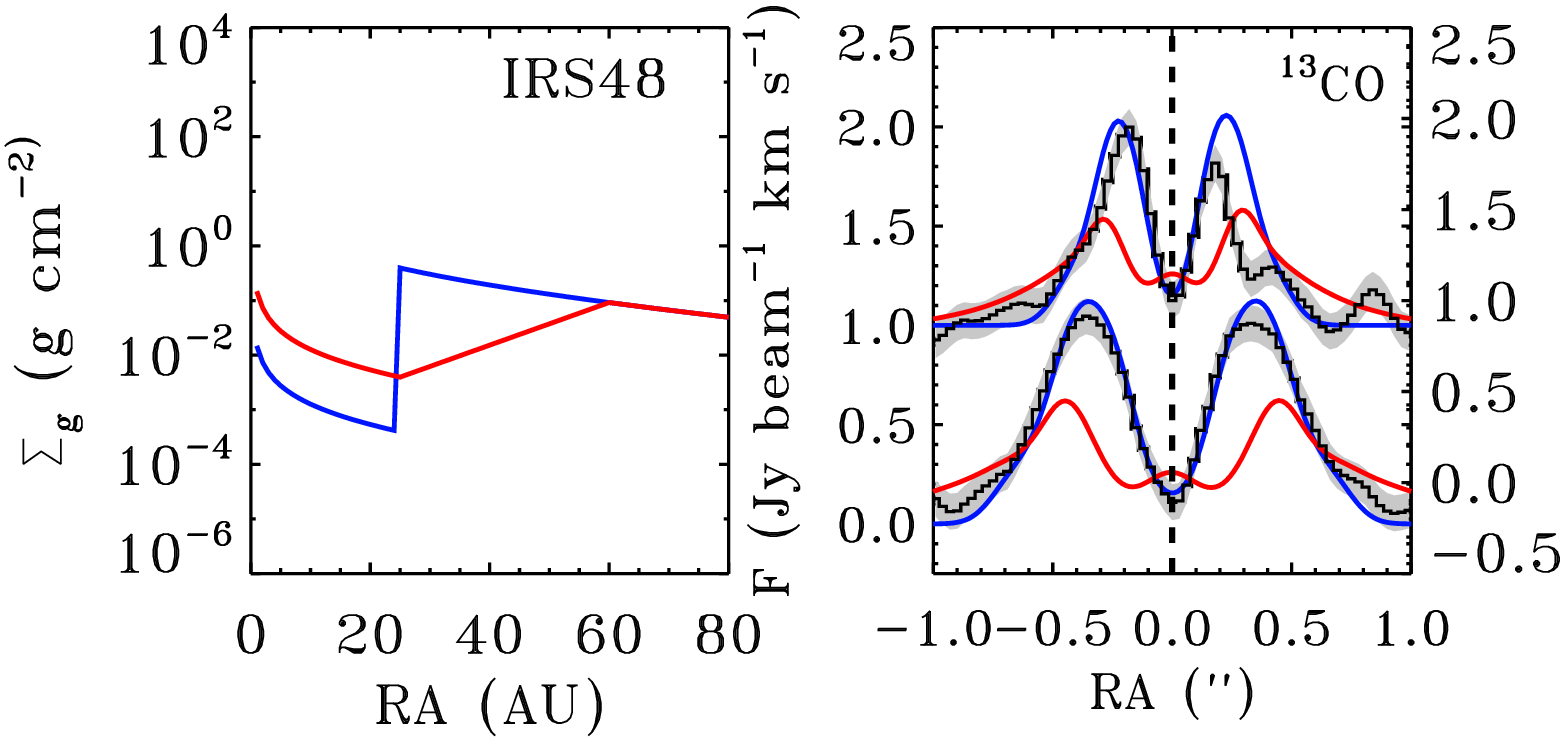}}
\caption{Comparison of the intensity cuts between the best-fit models (blue) and a gradual  model (red) as defined in Eq. \ref{eqn:slope}. The right panel gives the density profile. All disks except IRS48 can be fit equally well with this gradual increase profile.}
\label{fig:slopemodel}
\end{figure}

The structure with two gas density drops inside the cavity found in DoAr44 and SR21 (and previously in IRS 48, \citealt{Bruderer2014}) has been interpreted as the result of multiple planets at different orbits. An alternative explanation is a gradual drop or increasing surface density profile inside the cavity. To investigate this, we ran additional models and replaced the exponential power-law $\Sigma(r)$  by an increasing exponential law between $r_{\rm cavdust}$ and $r_{\rm cavgas}$, motivated by planet-disk interaction models, following
\begin{equation}
\label{eqn:slope}
\Sigma(r) = \Sigma(r_{\rm cav})\cdot e^{(r-r_{\rm cav2})/w}
,\end{equation}
where the width $w$ is given by
\begin{equation}
w=\frac{r_{\rm cav}-r_{\rm cav2}}{\ln (\Sigma(r_{\rm cav})/\delta_{\rm gas2}\Sigma(r_{\rm cav2})}
.\end{equation}
The width was chosen to connect $\Sigma(r)$ at $r_{\rm cavdust}$ and $r_{\rm cavgas}$, fitting $\delta_{\rm gas2}$, and no new parameters were introduced. This straight connection is further motivated by the shape of the gaps in Fig. 7 in \citet{deJuanOvelar2013} of planet-disk interaction models. Using the above relation and the derived cavity radii, the drop in density is derived again for the best-fitting model (see Fig. \ref{fig:slopemodel}). The best-fitting values are within a factor of 2 of the $\delta_{\rm gas2}$ values in the vertical drop model (Table \ref{tbl:physmodel}): we find 2$\cdot10^{-4}$, 10$^{-3}$, 2$\cdot10^{-4}$ , and 5$\cdot10^{-2}$ for HD135344B, SR21, DoAr44, and IRS48 respectively, although IRS48 is a poor fit compared to the double-drop model. However, the outer radius of the slope ($r_{\rm cav}$) of 60 AU has been constrained from the SED and VISIR image instead of from the millimeter continuum, as in the other disks, because of the asymmetric structure. This makes this approach rather uncertain. Another possibility is that a combination with other clearing mechanisms is responsible for the different drop shapes.

\subsection{Isotope-selective photodissociation}
\label{sct:photodiss}

The main process regulating the survival of CO in disks, photodissociation, does not equally affect different CO isotopologs \citep{vanDishoeckBlack1988}. $^{12}$CO becomes optically thick at low column densities and shields itself throughout the disk from the photodissociating flux. On the other hand, less abundant isotopologs, like C$^{18}$O, are not self-shielded until much deeper into the disk and continue to be photodissociated. This results in regions where C$^{18}$O is less abundant than predicted by a constant [$^{16}$O/$^{18}$O] ratio found in the interstellar medium. Accordingly, this leads to reduced C$^{18}$O line intensities. The importance of isotope-selective effects varies depending on disk parameters, dust properties, and the stellar far-ultraviolet field \citep{Miotello2014}. The UV field is calculated at each position in the DALI model taking into account the local dust density.

\begin{figure}[!ht]
\subfigure{\includegraphics[scale=0.35,trim=40 40 40 10]{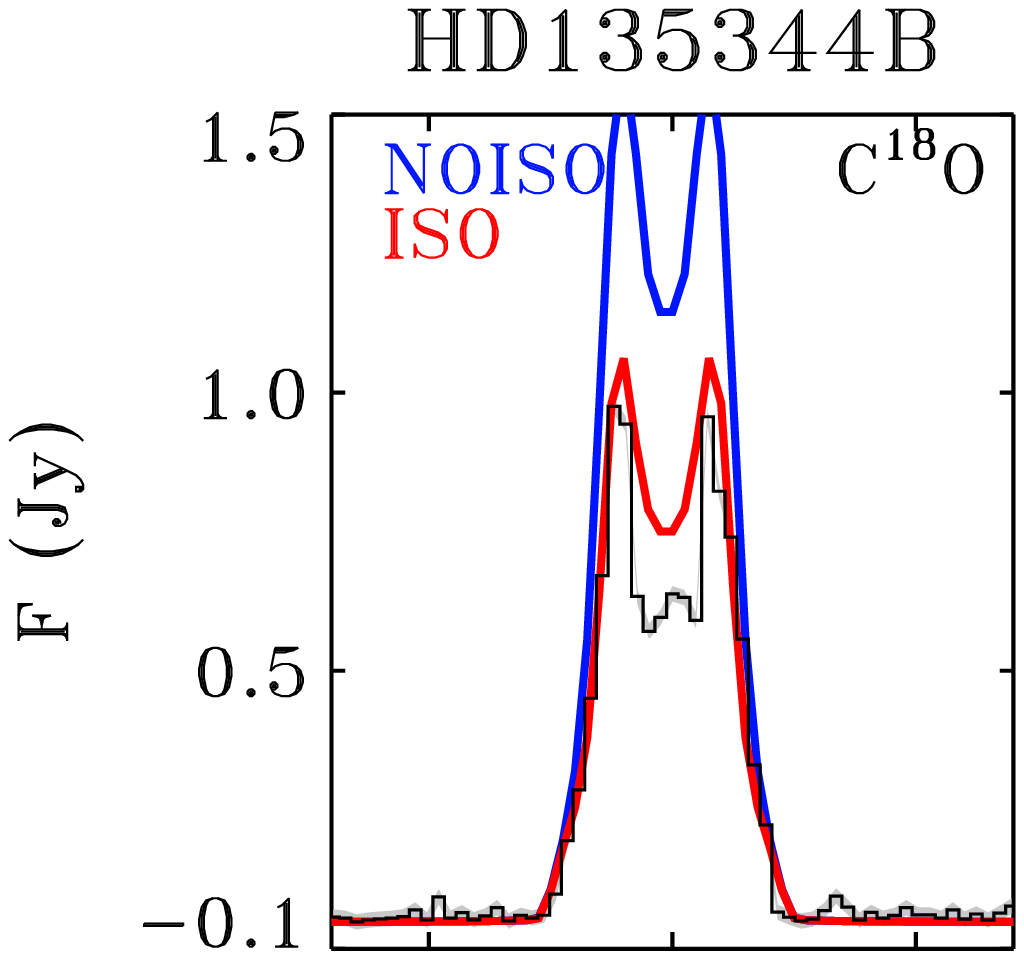}}
\subfigure{\includegraphics[scale=0.35,trim=40 40 40 10]{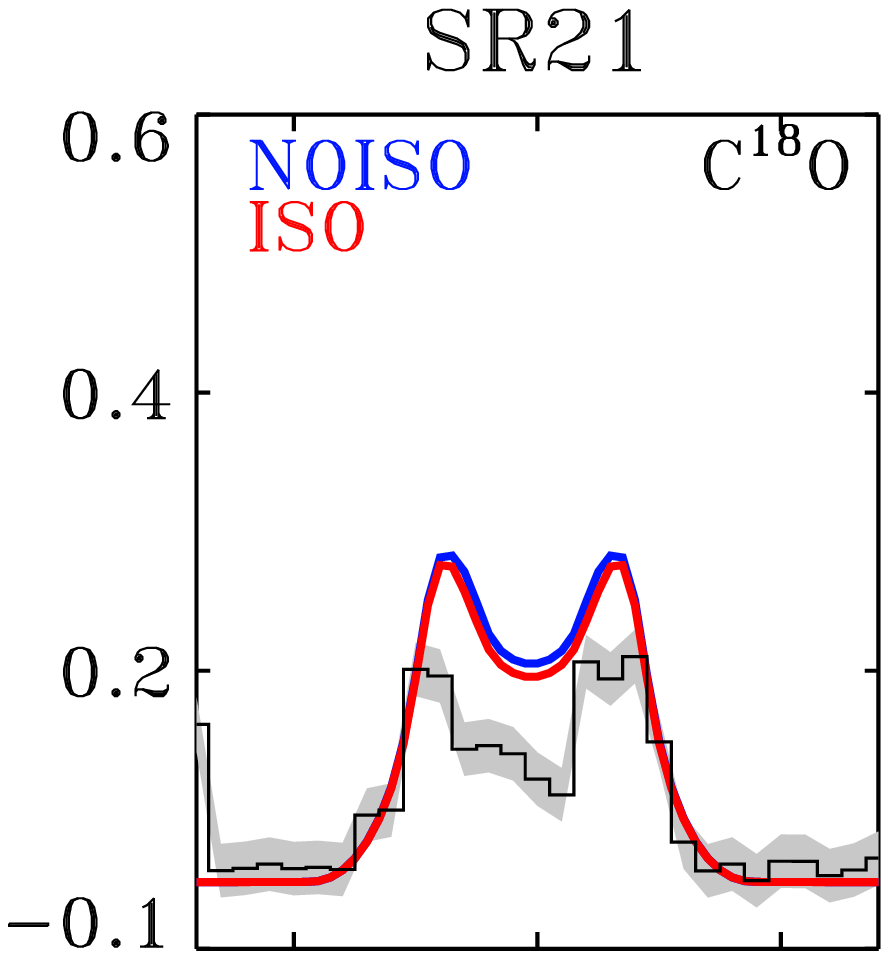}}\\
\subfigure{\includegraphics[scale=0.35,trim=40 20 40 40]{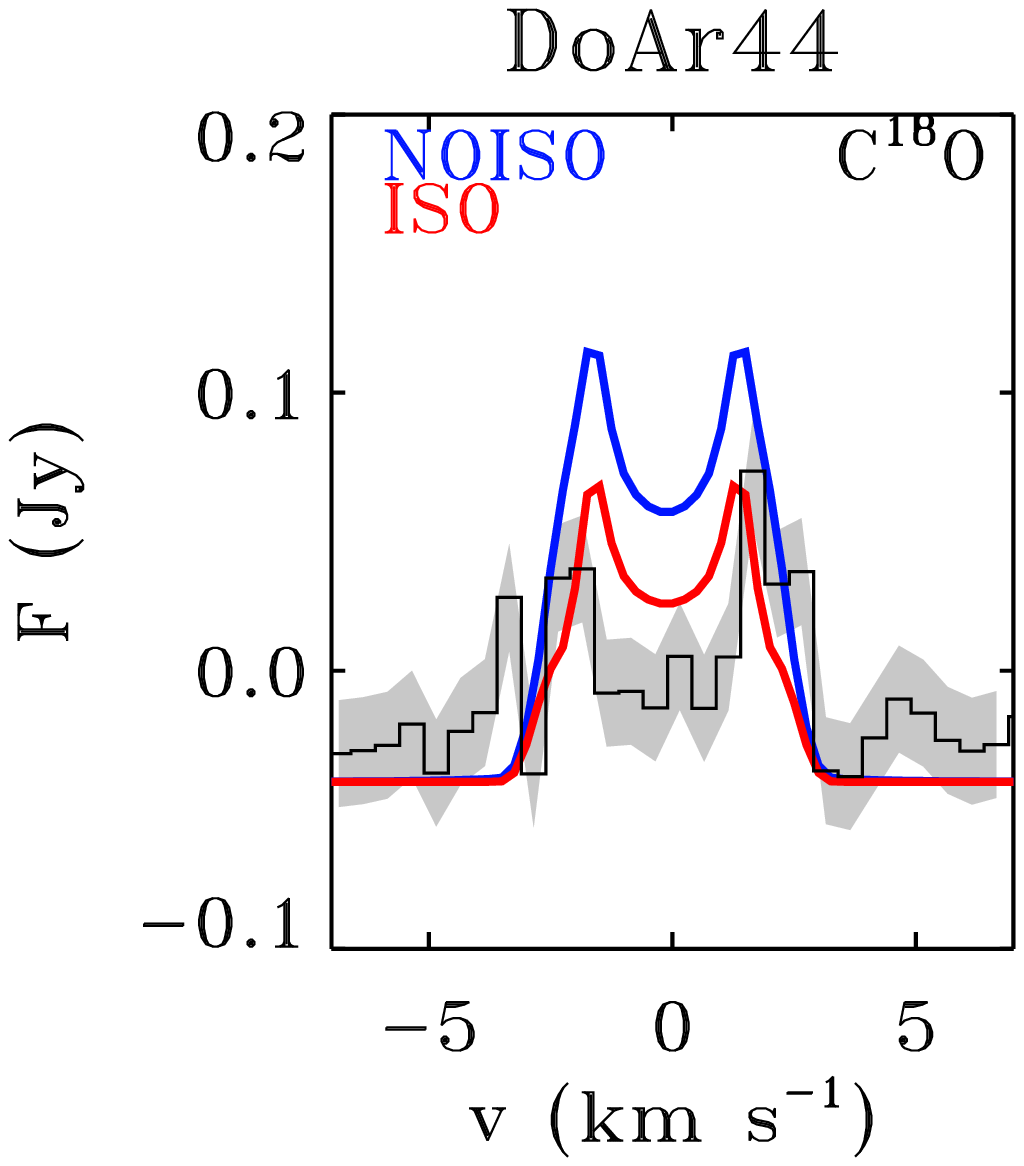}}
\subfigure{\includegraphics[scale=0.35,trim=40 20 40 40]{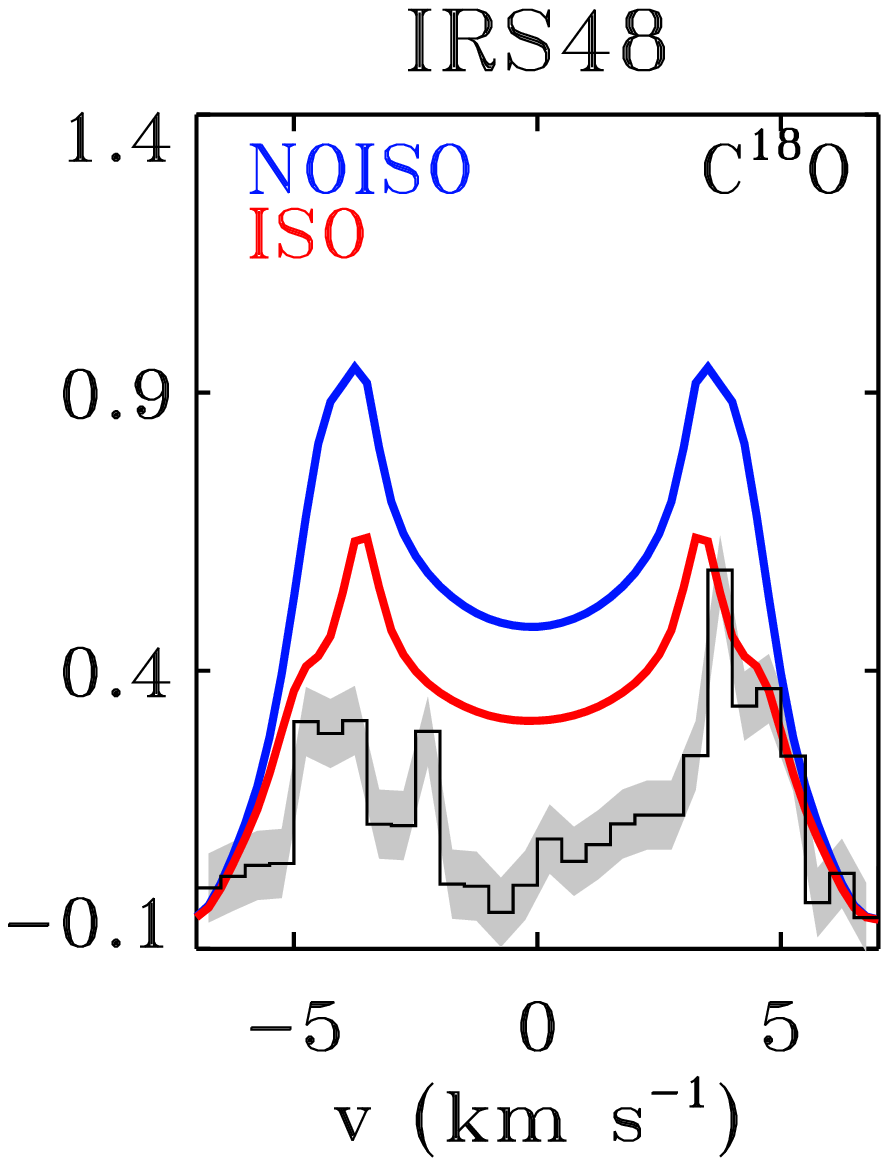}}
\caption{Comparison of the spectra of the C$^{18}$O emission with (red, ISO) and without (blue, NOISO) implementing isotope-selective photodissociation. HD135344B, SR21, and IRS48 differ by up to a factor of 2, while no difference is seen for DoAr44.}
\label{fig:isomodel}
\end{figure}

For this reason, isotope-selective photodissociation has been implemented in the modeling of all disks in our sample. Isotope-selective effects are substantial only for C$^{18}$O line emission and in the outer disk regions, where the bulk of the gas phase CO is located (Fig. \ref{fig:isomodel}). The effect on the outer disk emission is not detectable for SR21. On the other hand, IRS48 has an even lower disk mass but shows a significant difference. This may be related to the low gas-to-dust ratio in this disk. This shows that isotope-selective photodissociation depends on the combination of disk and stellar parameters and not only on the total disk gas mass.

Interestingly, isotope-selective photodissociation only affects the CO isotopolog emission in the outer disk (instead of inside the cavity), probably because of the lower CO column densities and the temperatures. In these conditions the isotope-selective photodissociation is more effective. A similar result was found in full disks, where the regions strongly affected by isotope-selective processes were located in the outer colder regions \citep{Miotello2014}. However, this is only true for conditions in these four disks. Possible effects in more extreme scenarios cannot be excluded.

\section{Discussion}
The main outcome of Fig. \ref{fig:intensitycuts} and the modeling is that all four transition disks have gas inside the cavity, with a gas cavity that is smaller than the dust cavity (for SR21 adopted from \citet{Pontoppidan2008}). The CO isotopologs confirm the main result from the $^{12}$CO data: gas is present inside the dust cavities and decreases less in density than the millimeter-dust. With the new optically thin isotopolog data, the density profiles are now much better constrained than those based on $^{12}$CO data alone \citep{vanderMarel2015-12co}.

Two other transition disks  have been sufficiently spatially resolved in CO and continuum to also confirm a gas cavity smaller than the dust cavity: HD142527 \citep{Fukagawa2013,SPerez2015} and J1604-2130 \citep{Zhang2014,vanderMarel2015-12co}. To date, no counterexample has been found for which the gas follows the same distribution as the dust. This hints at the exciting possibility that the origin of transition disks indeed lies in embedded planets that have cleared their orbit in the gas and trapped the millimeter-dust at the edge.  

An upper limit on any embedded companions has been derived for HD135344B using direct imaging in \citet{Vicente2011}: less than 1 brown dwarf mass at $>$37 AU radius. For SR21, companions with $q>0.01$ or $>10 M_{\rm Jup}$ are ruled out for 11-21 AU orbital radius according to near-infrared aperture masking interferometry \citep{Andrews2011}. For the other two disks no limits on companions are known. Unfortunately, the derived limits are outside the orbital range suggested by our gas cavity radii, assuming the companion orbital radius is close to the gas cavity radius. 

The difference between gas  and dust cavity radius can be compared with modeling results of planet-disk interaction by \citet{deJuanOvelar2013}. They show a relation between the observed dust cavity wall in the near-infrared (SPHERE-ZIMPOL predictions), tracing the small dust grains, and the millimeter-dust cavity wall as observed by ALMA. Simulations were run for different planet masses at different radii (Fig. 8 and Eq. 1 in de Juan Ovelar et al.). The simulations were performed assuming a viscosity of $\alpha\sim10^{-3}$ for a disk of 0.05 M$_{\odot}$. The ratio between the two radii $f(M_p)$ is found to follow
\begin{equation}
f(M_p) = c\cdot\left(\frac{M_p}{M_{\rm Jup}}\right)^{\gamma}
,\end{equation}
with $c\sim0.85$, and $\gamma\sim[-0.22,-0.18,-0.16]$ for planet orbital radius = [20, 40, 60] AU, respectively. Assuming that the small dust grains follow the gas, this relation can be directly applied to our cavity radii. The gas-to-dust cavity radii ratios for our targets are 0.75, 0.28, 0.5, and 0.42 for HD135344B, SR21, DoAr44, and IRS48, respectively, with gas cavity radii of 30, 7, 16, and 25 AU. Using the $\gamma=-0.22$ relation (closest to our gas cavity radii), planet masses are predicted to be 2, 11, and 25 $M_{\rm Jup}$ for HD135344B, DoAr44, and IRS48. For SR21, we extrapolate $\gamma$ for the 7 AU gas cavity radius to be -0.26, resulting in a planet mass of 71 Jupiter masses. The derived masses remain uncertain because of the fixed viscosity in the models: a lower viscosity would result in lower masses.

Another way to compare the observations with planet-disk interaction models is using the $\delta_{\rm gas2}$ drop value. The outcome of the gradual drop model is particularly interesting because
it better resembles the shape of the gap carved by a planet, as seen in planet-disk interaction simulations \citep{Crida2006, Zhu2011,deJuanOvelar2013,Fung2014}. Fung et al. (2014)\ have derived an analytical prescription based on the outcome of numerical simulations of the depth of the gap, which can set constraints on the mass of the planet  in combination with the viscosity in their Eq. 14:
\begin{equation}
\Sigma_{\rm gap}/\Sigma_0 = 4.7\times10^{-3}\left(\frac{q}{5\times10^{-3}}\right)^{-1}\left(\frac{\alpha}{10^{-2}}\right)^{1.26}\left(\frac{h/r}{0.05}\right)^{6.12}
,\end{equation}
with $q$ the mass ratio between planet and star, $\alpha$ the viscosity parameter, $h/r$ the scale height, and $\Sigma_{\rm gap}/\Sigma_0$ the drop in density, or $\delta_{\rm gas2}$. The equation is only valid for $q=10^{-4}-10^{-2}$. A similar relation was recently derived by \citet{Kanagawa2015}. The derived parameters can thus provide an estimate for the planet mass, assuming a certain viscosity value. The relation has a very strong dependence on $h/r$ because of the strong dependencies of the torque on the disk angular frequency. We note that the disks of these simulations are isothermal to make $h/r$ constant as a function of radius. Since the gas temperature has in reality a strong vertical gradient that is due to the UV heating, especially at the cavity wall, this makes Eq. 5 uncertain. On the other hand, these processes occur close to the midplane, and the isothermal approximation is not entirely incorrect. In applying the relation to our findings, $h/r$ is only marginally constrained by our models because of the degeneracies in SED modeling. As $h/r = c_s/v_k$, with the sound speed $c_s\propto\sqrt{T}$ and $v_k$ the Keplerian velocity, it can also be computed directly from the derived midplane temperatures. We find $h/r$ in our models at the gas cavity radius of 0.077, 0.063, 0.048, and 0.11 for HD135344B, SR21, DoAr44, and IRS48, respectively, which is generally not too different from the $h/r$ derived from our radiative transfer modeling. Using these values for $h/r$ in combination with the $\delta_{\rm gas2}$ values derived for the gradual drop models for our disks, planet masses can be derived for $\alpha$ between 10$^{-2}$ and 10$^{-4}$.  
For $\alpha=10^{-2}$, the $q$-values are $>0.01$ (except for DoAr44), which is beyond the range for which the analytical relation was derived. Higher $q$-values, implying substellar mass companions, would result in eccentric gaps and/or entire disruption of the disk, which is why the relation is no longer valid. For reasonable planet masses (up to 10 Jupiter masses) that are consistent with the upper limits for companions mentioned above, this implies low viscosity values between $\alpha\sim10^{-3}$ and 10$^{-4}$, much lower than found in the TW Hya and HD163296 disks based on turbulent broadening \citep{Hughes2011}.

The estimates derived here remain highly uncertain because of the uncertainties in our modeling and the assumptions in the numerical models: the relation was empirically derived based on the outcome of numerical simulations with several important limitations: the vertical structure is isothermal, accretion onto the planet is ignored, and the dust and gas are coupled, which is not true for a realistic physical disk. However, the $\delta_{\rm gas}$ or $\Sigma_{\rm gap}/\Sigma_0$, has been measured for the first time with an accuracy of better than an order of magnitude with these new observations. This parameter is inversely linear with $q$ or planet mass and sets a constraint on the properties of these potential embedded planets.

\section{Conclusions}
We have analyzed high spatial resolution ALMA submillimeter observations of $^{13}$CO and C$^{18}$O line emission from four transition disks using full physical-chemical modeling. Using a previously derived surface density model of the dust, based on the SED and millimeter continuum visibilities, a physical model of the gas and dust was derived for each of the disks. The structure and amount of gas inside the cavity is the main point of interest as it gives direct information about potential embedded planets. 
\begin{enumerate}
\item All four disks show a gas cavity that is up to twice as small in radius as the dust cavity. Two other examples are known from the literature.
\item All disks can be fit to a gas density model with one or two drops in the gas density inside the cavity.
\item The gas density inside the cavity drops by at least a factor of 1000 compared to the gas surface density profile of the outer disk.
\item An alternative model with a gradual increase of surface density with radius inside the cavity fits the data equally well for three of the four disks.
\item The derived values of the gas mass from the CO isotopologs are within a factor of a few compared to previously derived values from spatially resolved $^{12}$CO observations, submillimeter continuum, and a gas-to-dust ratio of 100. The isotopologs are crucial for the gas density profile inside the cavity, however.
\item The depth of the gas density drop indicates that the viscosities in these disks are low for reasonable companion masses.
\end{enumerate}
These spatially resolved isotopolog data point to embedded planets as by far the most likely explanation.

  \begin{acknowledgements}
  The authors would like to thank P. Pinilla for useful discussions.
  Astrochemistry in Leiden is supported by the Netherlands
  Research School for Astronomy (NOVA), by a Royal Netherlands Academy
  of Arts and Sciences (KNAW) professor prize, and by the European
  Union A-ERC grant 291141 CHEMPLAN. This paper makes use of the
  following ALMA data: ADS/JAO.ALMA/2012.1.00158.S and ADS/JAO.ALMA/2013.1.00100.S. ALMA is  a partnership of ESO (representing its member states), NSF (USA) and
  NINS (Japan), together with NRC (Canada) and NSC and ASIAA (Taiwan),
  in cooperation with the Republic of Chile. The Joint ALMA
  Observatory is operated by ESO, AUI/NRAO and NAOJ.  
  \end{acknowledgements}

\bibliographystyle{aa}

\appendix
\onecolumn
\section{Channel maps}
\label{sct:channels}
In this section we present the $^{13}$CO channel maps for each observed target.
\begin{figure*}[!ht]
\begin{center}
\subfigure{\includegraphics[scale=0.9]{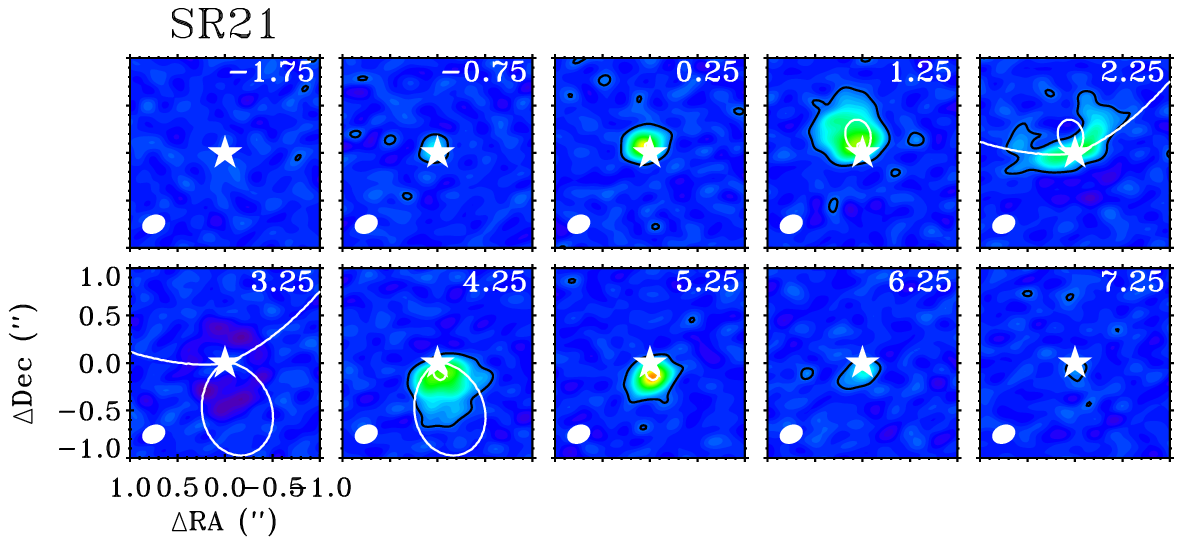}}\\
\subfigure{\includegraphics[scale=0.9]{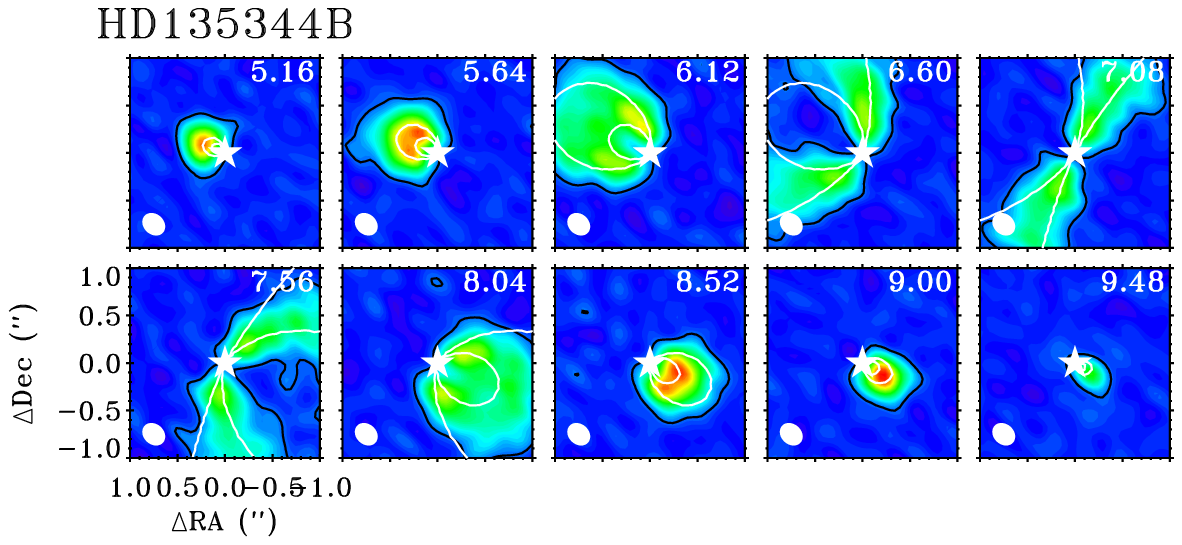}}\\
\subfigure{\includegraphics[scale=0.9]{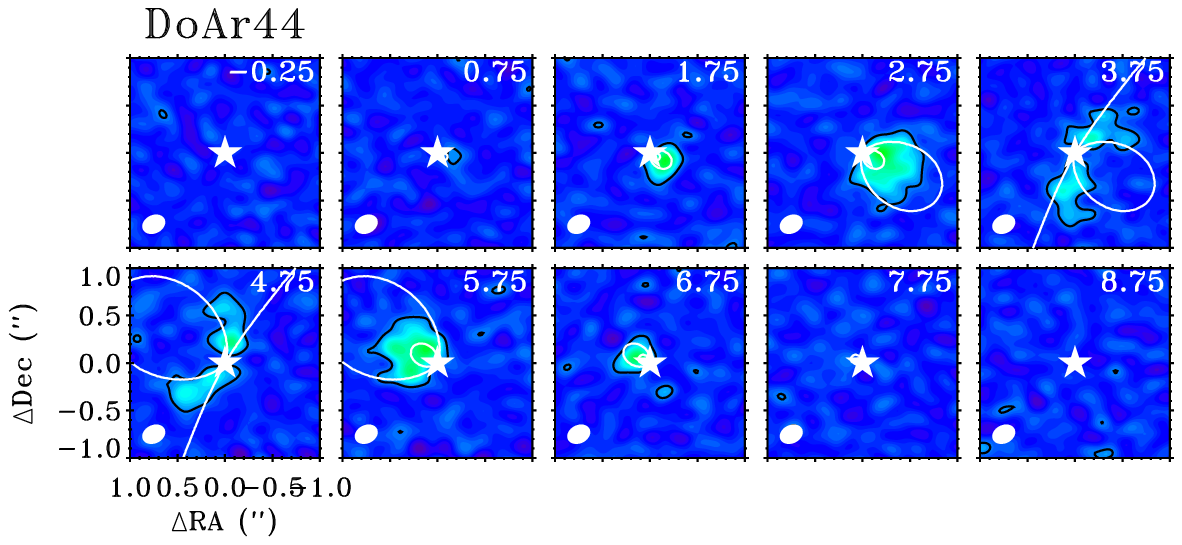}}\\
\subfigure{\includegraphics[scale=0.9]{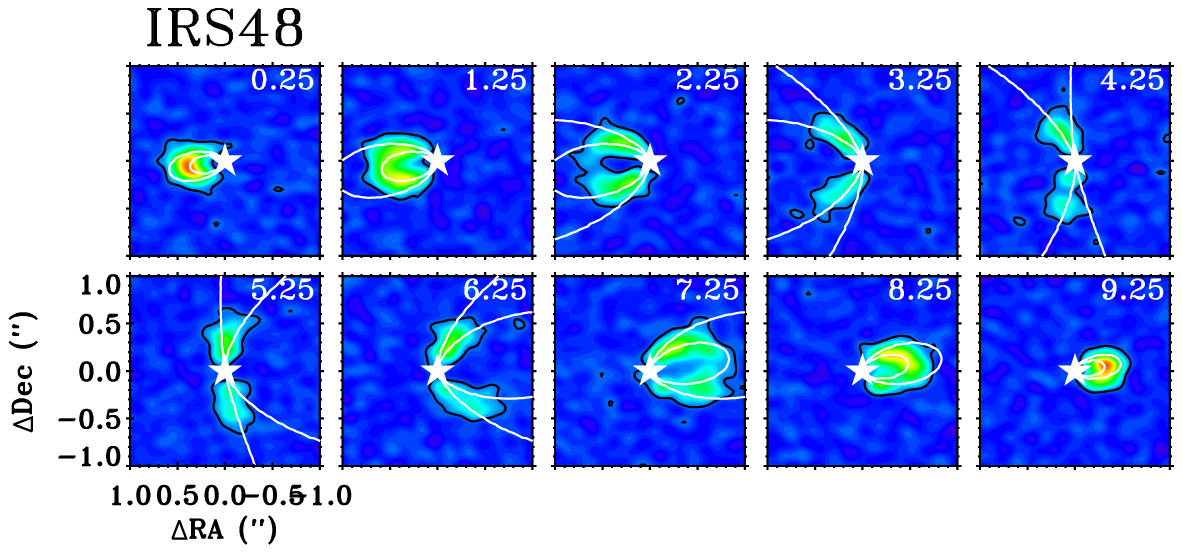}}
\end{center}
\caption{$^{13}$CO channel maps for each observed target. Overlaid in white contours are the Keplerian velocity profiles for the derived inclination and the given stellar mass.}
\label{fig:channelmap}
\end{figure*}

\section{Dust model DoAr44}
\label{sct:doar44model}
Unlike the other disks, the dust surface density profile of DoAr44 was not yet constrained by ALMA data from previous papers. Using the same approach as in  \citet{vanderMarel2015-12co}, a model was found by fitting the SED, the 345 GHz continuum visibility curve, and the 345 GHz continuum intensity cuts. The cavity size is 32 AU, similar to the previous result by \citet{Andrews2011}. The depth of the dust density inside the cavity was constrained by varying the $\delta_{\rm dustcav}$ between $r_{\rm gap}$ and $r_{\rm cavdust}$. It was found that the dust density drops by at least a factor 10$^3$, similar to the other disks.

\begin{figure*}[!ht]
\includegraphics[scale=0.5]{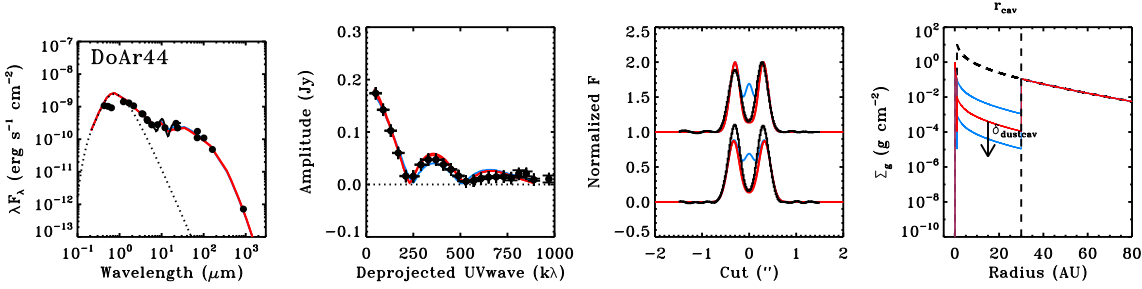}\\
\caption{Modeling results and observations of DoAr44 of the dust surface density, comparing $\delta{\rm dustcav}$ ranging between 10$^{-2}$ and 10$^{-4}$ as indicated in the right panel. The observations are plotted in black. {\bf Left:} Spectral energy distribution. {\bf Middle left:} Amplitude of the 345 GHz continuum visibility for the deprojected baselines.  The null line is indicated with a dashed line. {\bf Middle right:} Normalized intensity cuts through the major (bottom) and minor (top) axis of the 345 GHz continuum image. The model images are convolved with the same beam as the ALMA observations. {\bf Right:} The dust surface density profile. Indicated are the $\delta_{\rm dust}$ , the drop in density to fit the inner disk through the near-infrared emission, and $\delta_{\rm dustcav}$ , the minimum drop in dust density inside the cavity needed to fit the observations.}
\label{fig:dustcavmap}
\end{figure*}

\section{Additional models}
Figure \ref{fig:changecavitysize} shows the modeling results for $^{13}$CO for the baseline model (Table \ref{tbl:physmodel}) for different gas cavity sizes. The plots show that the gas cavity radius is determined to within an uncertainty of 5 AU.

\label{sct:additionalmodels}
\begin{figure*}[!ht]
\begin{center}
\subfigure{\includegraphics[scale=0.6]{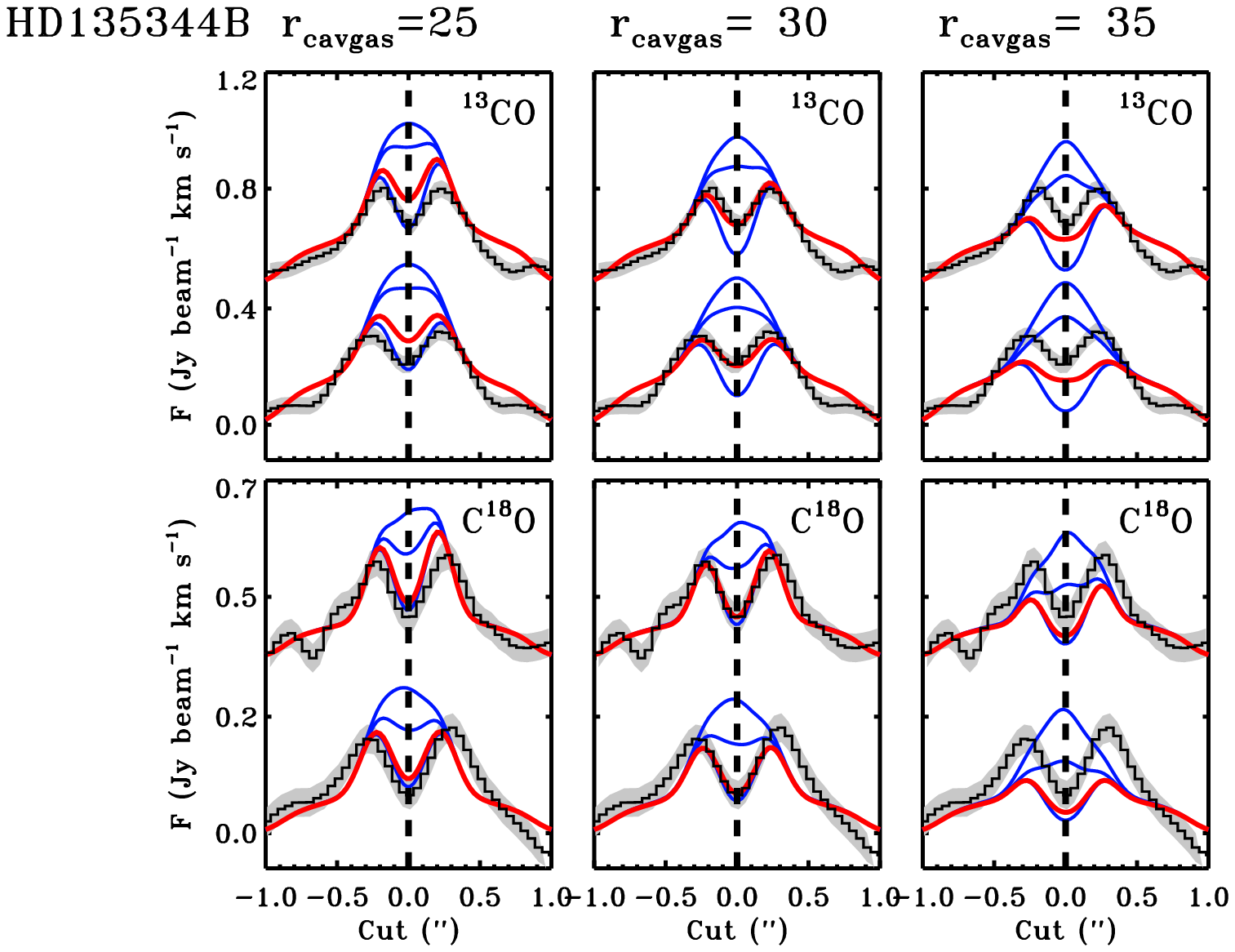}}\\
\captcont{Modeling results for different cavity sizes for three of the four sources. The plots show the $^{13}$CO intensity cuts of the data (black) and the models (colors) for different density drops $\delta_{\rm gas2}$. The central panel uses the gas cavity size $r_{\rm cavgas}$ of the final model (see Table \ref{tbl:physmodel}), the left panel the results for a 5 AU smaller gas cavity, and the right panel the results for a 5 AU larger cavity. For SR21 the 7 AU radius remains unresolved in the ALMA data so the radius is not explored. The plots reveal that the gas cavity radius is determined to within 5 AU uncertainty.}
\end{center}
\label{fig:changecavitysize}
\end{figure*}

\begin{figure*}
\begin{center}
\subfigure{\includegraphics[scale=0.6]{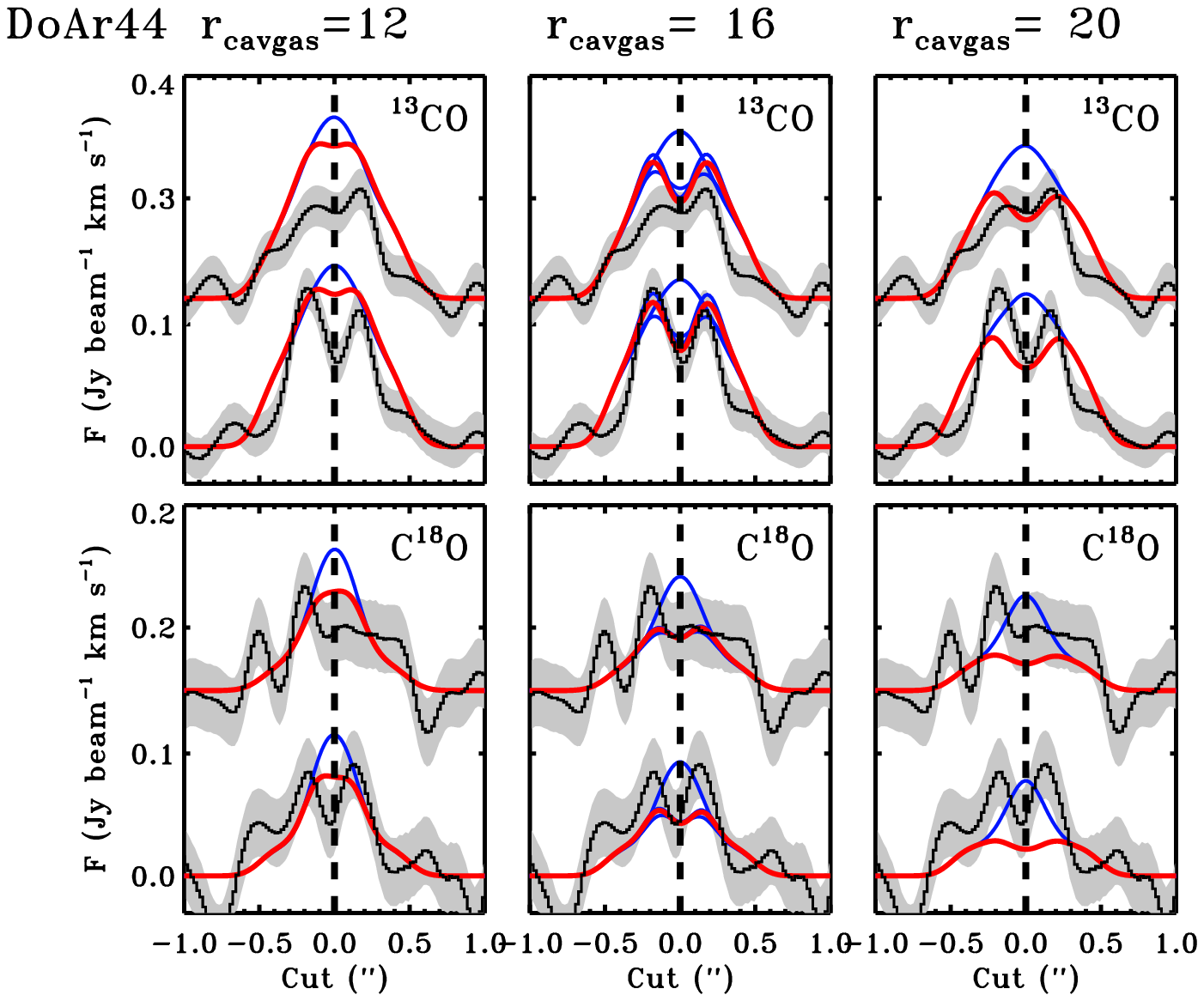}}\\
\subfigure{\includegraphics[scale=0.6]{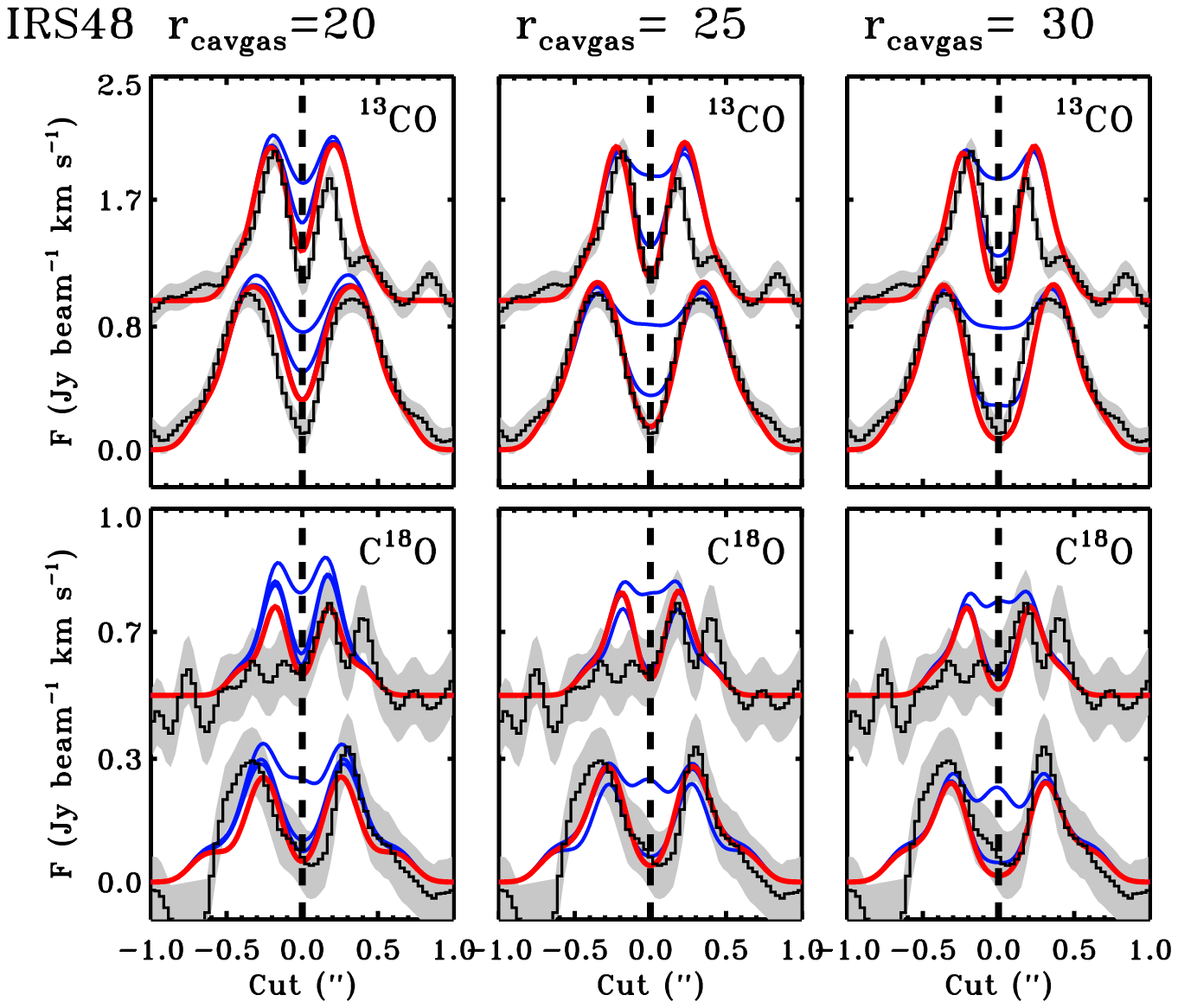}}
\end{center}
\caption{Continued.}
\end{figure*}
\end{document}